\documentclass[fleqn,usenatbib]{mnras}

\usepackage[T1]{fontenc}

\DeclareRobustCommand{\VAN}[3]{#2}
\let\VANthebibliography\thebibliography
\def\thebibliography{\DeclareRobustCommand{\VAN}[3]{##3}\VANthebibliography}

\usepackage{graphicx}	
\usepackage{amsmath}	
\usepackage{amssymb}	

\usepackage{rotfloat}
\usepackage{caption}
\usepackage[normalem]{ulem}

\usepackage{newtxtext,newtxmath}

\newcommand{\kms}{km\,s$^{-1}$}

\title[Triangulum II]{Stellar kinematics of dwarf galaxies from multi-epoch spectroscopy: application to Triangulum II 
\thanks{New observations reported here were obtained at the MMT Observatory, a joint facility of the Smithsonian Institution and the University of Arizona.  Some of the data presented herein were obtained at the W. M. Keck Observatory, which is operated as a scientific partnership among the California Institute of Technology, the University of California and the National Aeronautics and Space Administration. The Observatory was made possible by the generous financial support of the W. M. Keck Foundation.}
}

\author[R. Buttry et al.]{
Rachel Buttry$^{1}$\thanks{E-mail: rbuttry@andrew.cmu.edu (RB)},
Andrew B. Pace$^{1}$,
Sergey E. Koposov$^{2,3,1}$,
Matthew G. Walker$^{1}$,
Nelson Caldwell$^{4}$,
\newauthor
Evan N. Kirby$^{5}$,
Nicolas F. Martin$^{6,7}$, 
Mario Mateo$^{8}$,
Edward W. Olszewski$^{9}$,
Else Starkenburg$^{10}$,
\newauthor
Carles Badenes$^{11}$, and
Christine Mazzola Daher$^{11}$
\\
$^{1}$McWilliams Center for Cosmology, Carnegie Mellon University, 5000 Forbes Ave, Pittsburgh, PA 15213, USA \\
$^{2}$Institute for Astronomy, University of Edinburgh, Royal Observatory, Blackford Hill, Edinburgh EH9 3HJ, UK\\
$^{3}$Institute of Astronomy, University of Cambridge, Madingley Rd, Cambridge, CB3 0HA, UK\\
$^{4}$Center for Astrophysics $|$ Harvard \& Smithsonian, 60 Garden Street, Cambridge, MA 02138, USA\\
$^{5}$California Institute of Technology, 1200 E. California Blvd., MC 249-17, Pasadena, CA 91125, USA \\
$^{6}$Observatoire astronomique de Strasbourg, Universit\'e de Strasbourg, CNRS, UMR 7550, 11 rue de l'Universit\'e, F-67000 Strasbourg, France \\
$^{7}$Max-Planck-Institut f\"{u}r Astronomie, K\"{o}nigstuhl 17, D-69117 Heidelberg, Germany\\
$^{8}$Department of Astronomy, University of Michigan, Ann Arbor, MI 48109, USA\\
$^{9}$Steward Observatory, The University of Arizona, 933 N. Cherry Avenue, Tucson, AZ 85721, USA\\
$^{10}$Kapteyn Astronomical Institute, University of Groningen,Postbus 800, 9700 AV, Groningen, the Netherlands\\
$^{11}$Department of Physics and Astronomy and Pittsburgh Particle Physics, Astrophysics and Cosmology Center (PITT PACC),\\
University of Pittsburgh, 3941 O‘Hara Street, Pittsburgh, PA 15260, USA\\
}

\date{Accepted XXX. Received YYY; in original form ZZZ}

\pubyear{2021}

\begin{document}
\label{firstpage}
\pagerange{\pageref{firstpage}--\pageref{lastpage}}
\maketitle

\begin{abstract}
We present new MMT/Hectochelle spectroscopic measurements for 257 stars observed along the line of sight to the ultra-faint dwarf galaxy Triangulum II.  Combining results from previous Keck/DEIMOS spectroscopy, we obtain a sample that includes 16 likely members of Triangulum II, with up to 10 independent redshift measurements per star.  To this multi-epoch kinematic data set we apply methodology that we develop in order to infer binary orbital parameters from sparsely sampled radial velocity curves with as few as two epochs.  For a previously identified (spatially unresolved) binary system in Tri~II, we infer an orbital solution with period $296.0_{-3.3}^{+3.8} \rm~ days$ , semi-major axis $1.12^{+0.41}_{-0.24}\rm~AU$, and systemic velocity $ -380.0 \pm 1.7 \rm~km ~s^{-1}$ that we then use in the analysis of Tri~II's internal kinematics.  Despite this improvement in the modeling of binary star systems, the current data remain insufficient to resolve the velocity dispersion of Triangulum II.  We instead find a 95\% confidence upper limit of $\sigma_{v} \lesssim 3.4 \rm ~km~s^{-1}$. 
\end{abstract}




\section{Introduction}

Dwarf galaxies are of great importance for astrophysics.  From a galaxy formation perspective, dwarf galaxies are among the oldest and least chemically evolved objects \citep{mateo98,tolstoy09,mcconnachie12}. From a dark matter perspective, they include the most dark matter dominated systems known, with published dynamical mass-to-light ratios reaching as high as $10^{4}$ in solar units  \citep[and references therein]{simon2019}.  In this vein, dwarf galaxies are believed to be key components in unpacking the mystery of dark matter, as they probe the small-scale structure (< 1 Mpc) regime of $\Lambda$CDM cosmology \citep{bullock2017}.

In order to place dwarf galaxies into their proper cosmological context, we must obtain accurate estimates of their dark matter content.  The simplest dynamical mass estimators, based on the assumption of dynamic equilibrium, are functions of the effective radius and line-of-sight velocity dispersion measured for the stellar component \citep[e.g.,][]{ illingworth76,walker09,wolf10,errani2018}.  However, measurements of the stellar velocity dispersion can be challenging.  One reason is the small number and low luminosities of stellar tracers in especially the `ultrafaint' dwarf galaxies.  Another challenge is the existence of unresolved binary stars, whose orbital motions add a time-dependent component to the velocities measured for individual stars, and---if unaccounted for---can thereby inflate measurements of dwarf galaxy velocity dispersions.  Binary orbital motions alone can generate apparent velocity dispersions of a few km s$^{-1}$ \citep{mcconnachie2010}.  While this effect is negligible for the more luminous dwarf spheroidals, which have intrinsic velocity dispersions of $\sim 10$ km s$^{-1}$ \citep{Olszewski1996}, it can potentially contribute significantly to the $\la 3$ km s$^{-1}$ dispersions observed for the least luminous galaxies \citep{mcconnachie2010, Minor2010ApJ...721.1142M}.  

Various strategies have been used to account for the effect of binary stars on velocity dispersions and the dynamical masses derived therefrom.  When multi-epoch spectroscopy is available, one can identify probable binary systems via their observed accelerations \citep[e.g.,][]{Olszewski1996, martinez2011, koposov11, Minor2019MNRAS.487.2961M}; indeed modeling of multi-epoch spectroscopic data sets for luminous dwarf spheroidals suggests typical binary fractions near $\sim 50\%$ \citep{minor2013, spencer2017, spencer2018}, consistent with studies of Galactic binaries that indicate relatively high multiplicity fractions at low metallicity \citep{badenes18}.  In some cases, the removal of suspected binary stars has a significant impact on the measured velocity dispersion \citep[e.g.,][]{kirby2017,venn2017}.  

The Triangulum II (Tri~II) ultra-faint dwarf galaxy provides an interesting case study.  The original kinematic study of Tri~II, based on single-epoch spectroscopy of six member stars, measured a velocity dispersion of $5.1^{+4.0}_{-1.4} \rm~ km~s^{-1}$ , suggesting a dynamical mass-to-light ratio of $3600_{-2100}^{+3500}$ in solar units, and an extremely high dark matter density of $4.8_{-3.5}^{+8.1} \rm ~ M_{\odot}/\mathrm{pc}^3$ \citep{kirby2015}.  An independent study by \cite{martin2016} obtained a spectroscopic sample of 13 member stars, finding complicated kinematics in which a central velocity dispersion of $\sigma_v = 4.4 ^{+2.8}_{-2.0}\rm~km~s^{-1}$ gives way to a larger value of $14.1_{-4.2}^{+5.8}$ km s$^{-1}$ at a large radius.  Both studies found evidence for nonzero metallicity dispersion, supporting the conclusion that Tri~II is a dwarf galaxy embedded in a massive dark matter halo, and not a self-gravitating star cluster.  

However, follow-up spectroscopy soon provided a time domain and revealed the presence of at least one star with significant velocity variability.  From high-resolution spectra obtained primarily to analyze chemical abundances, \cite{venn2017} measured a change in velocity for one star ({\it Star46}) of $\sim 25$ km s$^{-1}$ with respect to the initial epoch measured by \citet{martin2016}.  \citet{kirby2017} added additional epochs for this star, independently confirming its velocity variability and finding that, when they excluded the likely binary from their analysis, the velocity dispersion was unresolved.  These circumstances leave the case for a dominant dark matter halo in Tri~II resting on the indirect argument provided by its metallicity spread \citep{venn2017}.  

Here, we add to the saga of Tri~II in two ways.  First, we present new spectroscopic data acquired with the Hectochelle spectrograph at the 6.5-m MMT. Second, we combine with the previously published spectroscopic data in order to obtain a multi-epoch data set that then lets us model the orbital parameters of the likely binary star.  Our orbital solution includes an inference for the binary system's center-of-mass motion, allowing us properly to include this star in our analysis of Tri~II's stellar kinematics. 

To date, only one star system within a dwarf spheroidal galaxy has a full orbital solution, based on 34 independent velocity measurements taken over a two-year baseline \citep{koch2014}.  Here, we develop methodology for inferring orbital solutions with as few as two velocity epochs.  The problem of finding orbital parameters for a binary system given a small number of radial velocity (RV) measurements has been undertaken previously by \cite{price-whelan2017} to create \textit{the JOKER}. Like \textit{the JOKER}, the binary model we present in this paper takes the approach of performing rejection sampling with likelihood function marginalized over some orbital parameters. However, our method has the added modifications such as the marginalization over inclination rather than semi-amplitude (allowing for the calculation of semi-major axis), the ability to take non-trivial priors over binary parameters, and using parameter samples for hierarchical models of binary populations.  

In the next section we discuss the MMT and Keck catalogues used in this analysis as well as the calculation of a zero-point correction between the two instruments. In section \ref{sec:methods} we present our methodology for modeling of binary and non-binary star systems as well as the galaxy kinematics. We then detail the resulting orbital parameter for the Tri~II binary system and the overall Tri~II kinematics in section \ref{sec:results}. Lastly, we discuss the findings from our work in Section \ref{sec:conclusions} and we suggest a hierarchical model building off the methods used.

\section{Data}\label{sec:data}

\subsection{MMT Hectochelle}\label{sec:MMT}
During Dec. 2015 and Oct. - Nov. 2016, we acquired new spectra of stars in Tri~II using the Hectochelle spectrograph \citep{szentgyorgyi2011} at the 6.5-m MMT Observatory on Mt. Hopkins, Arizona.  Hectochelle deploys up to 240 optical fibers, each with aperture $ 1.5$ arcsec, over a field of diameter $1^{\circ}$.  We observed using the `RV31' filter, isolating the wavelength range of $5150-5300$ at resolution $\mathcal{R}\approx 34,000$.  

We observed five different Hectochelle fiber configurations, each centered on the published center of Tri~II, allowing us to observe up to $\sim 500$ unique targets, with many stars included in multiple targeting configurations.  

To clean the overwhelming contamination from foreground Milky Way stars, we rely on narrow-band, metallicity sensitive CaHK observations of Tri~II. The observations follow a similar strategy and goal as presented by \cite{starkenburg2017} in the Pristine survey but correspond to a single CFHT MegaCam field centered on Tri II. The field covers the full extent of Tri~II and the integration amount to 4,800 s in total from the CaHK images. Additional observations were also obtained in the MegaCam g and i bands (4,140 and 4,050 s, respectively) to benefit from the full depth of the CaHK photometry is deeper than the Pan-STARRS1 data that were used to discover Tri II \citep{Laevens2015}. All were observed in service mode by the CFHT observing staff between 2015-07-18 and 2016-02-13. After pre-processing of the images with ELIXIR by CFHT \citep[de-biasing, flat-fielding, and de-trending;][]{Magnier2004}, the images were astrometrically calibrated, stacked, and processed for photometry with the CASU pipeline \citep{irwin2001}, as described in detail by \cite{starkenburg2017}. The broadband $g$ and $I$ photometry is calibrated onto the Pan-STARRS1 $g_\mathrm{P1}$ and $i_\mathrm{P1}$ bands and we directly use the Pan-STARRS1 photometry at the bright end, where the CFHT observations saturate. Finally, the photometric catalog is de-reddened using the Schlegel, Finkbeiner and Davis dust maps.

\begin{figure}
    \centering
    \includegraphics[width=\columnwidth]{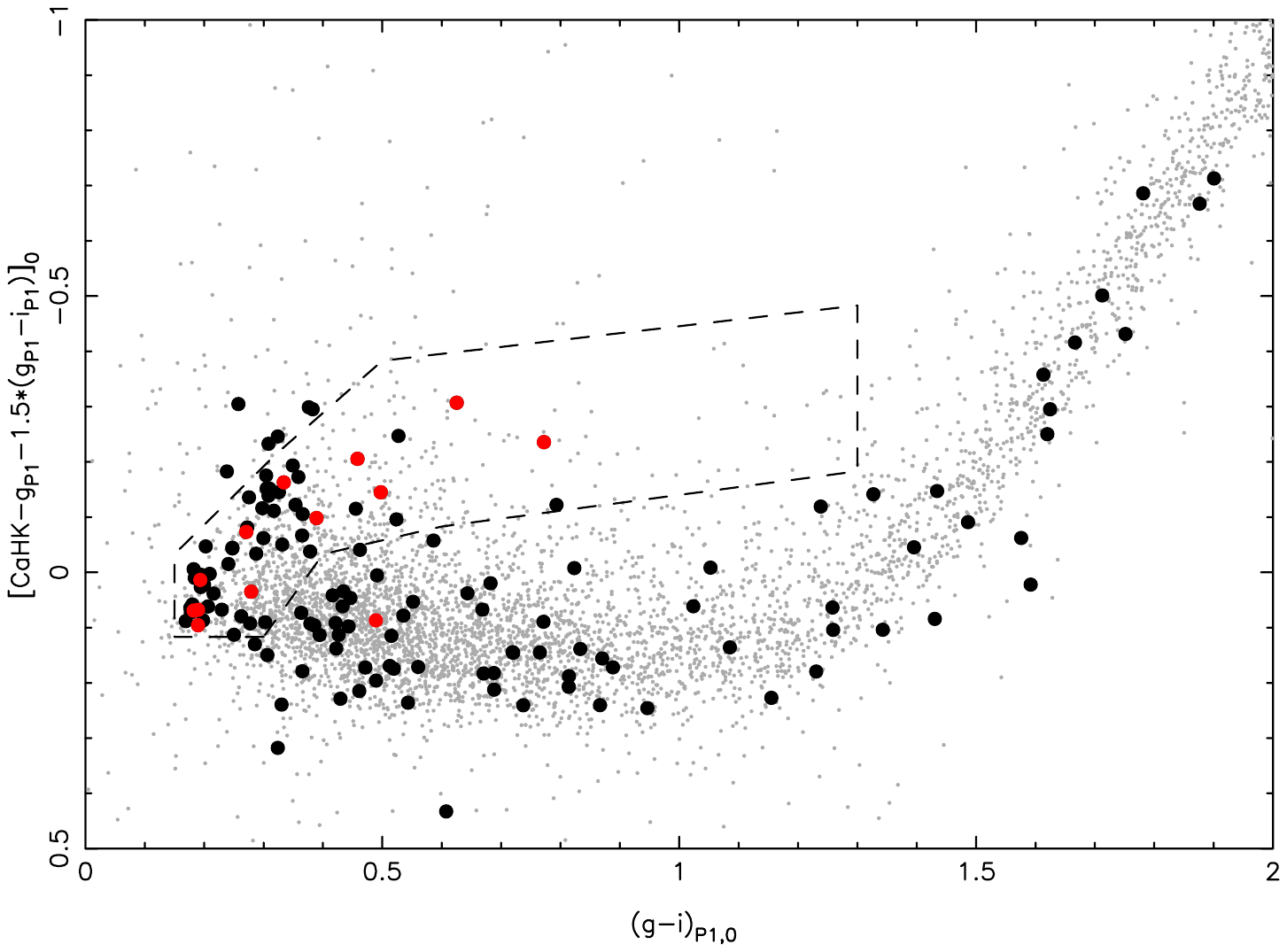}
    \caption{(CaHK, g, i) color-color space for Tri II. All stars in the MegaCam photometry are shown as small grey dots and follow a stellar locus that is produced by Milky Way, metal-rich stars. In this space, metal-poor stars are above this locus. Large black symbols correspond to stars within 4' of the center of Tri II, with radial-velocity-selected, likely members shown in red. One of those stars is shown here to be metal-rich and unlikely to be a true members. Our selection for Hectoshell gives the highest priority to stars within the dashed polygon that provides a loose selection of stars away from the metal-rich foreground contamination.}
    \label{fig:pristine}
\end{figure}

Similarly to what was done for the Pristine Inner Galaxy Survey \citep{arentsen2020} the large number of science fibers available on Hectochelle allowed us to perform a broad selection of targets in the part of the CaHK color-color diagram that contains metal-poor stars. This allowed us to bypass any potential calibration issue and the use of Pan-STARRS1 broad-band magnitudes instead of the SDSS ones we relied on in \cite{starkenburg2017}. In particular, we selected stars based on their color-magnitude diagram location so they broadly follow an old and metal-poor isochrone. In the color-color space using CaHK that is presented in Figure \ref{fig:pristine} and used by the Pristine survey, we loosely select stars from the metal-poor region, using known Tri II members as a guideline to isolate other stars with similar properties. Once fibers are assigned to these high-priority stars, we fill the rest of the fibers with random CMD-selected stars, irrespective of their CaHK information. 

We processed  all raw Hectochelle spectra using the CfA pipeline (\texttt{HSRED}~v2.1\footnote{\url{https://bitbucket.org/saotdc/hsred/} }). Following the procedure described in detail by \citet{walker15draco}, we then analyzed each individual spectrum by fitting a model based on a library of synthetic template spectra that span a regular grid in effective temperature, surface gravity and [Fe/H] metallicity.  In addition to the stellar-atmospheric parameters, we fit for line-of-sight velocity as well as several free parameters that specify the continuum shape and correct for wavelength-dependent velocity shifts.

With respect to the procedure documented by \citet{walker15draco}, for present purposes we update our estimation of systematic errors associated with line-of-sight velocity and metallicity.  For this task we use our entire catalog of MMT/Hectochelle observations of dwarf galaxies and globular clusters, including observations spanning the years 2005 to 2020.  This sample includes 12517 independent observations of 7906 unique stars, including 2501 stars with up to 13 individual measurements.
We model the pair-wise velocity and metallicity differences as a mixture of a Gaussian with an outlier model (see Section 4.1 of \citealt{li2019} and Section 2.2 of~\citealt{pace2021}).The final uncertainty ($\sigma_{v, \rm calib}$) is treated as a systematic error ($\sigma_{v, \, {\rm systematic}}$) plus a scaling parameter ($k_v$) and $\sigma_{v, \rm calib}^2 = \sigma_{v, \, {\rm systematic}}^2 + (k_v \sigma_{v, {\rm mcmc}})^2$. We find $k_v=1.03 \pm 0.02$ and $\sigma_{v, \, {\rm systematic}}=0.35\pm 0.02$ \kms  for the velocity systematic errors, and $k_{\rm [Fe/H]}=1.33\pm0.01$ and $\sigma_{{\rm [Fe/H]}, \, {\rm systematic}}=0.0\pm0.01$ for metallicity systematic errors.

\subsection{Keck DEIMOS}\label{sec:Keck}
Our sample also includes spectroscopy obtained with the Deep Extragalactic Imaging Multi-Object Spectrograph \citep[DEIMOS;][]{faber2003}.  First, \citet{kirby2015,kirby2017} observed six slitmasks in 2015 and 2016.  They used the 1200G grating, which achieves a spectral resolution of $R \sim 7000$ at 8500~\AA, in the spectral vicinity of the Ca$\,${\sc ii} infrared triplet.  Second, \citet{martin2016} observed two DEIMOS slitmasks with a similar spectral configuration as \citeauthor{kirby2015}  

\citet{kirby2015} selected stars using Keck/LRIS \citep{oke1995} photometry.  They chose a generous selection region in the CMD around the red giant branch as defined by the ridgeline of the globular cluster M92.  \citet{martin2016} used a similar selection technique with photometry \citep{Laevens2015} from the Large Binocular Camera.  In general, the field of Tri~II is sparse enough that most candidate member stars in the field of the slitmask could be observed.  As a result, the samples have little selection bias due to color (or stellar age or metallicity).  \citet{kirby2017} were mainly interested in quantifying radial velocity variability, not in finding new members.  As a result, they designed their slitmasks to target stars already identified as members by \citet{kirby2015} and \citet{martin2016}.

Velocities were measured in slightly different ways.  \citet{kirby2015,kirby2017} reduced the spectra with custom modifications to the spec2d data reduction pipeline \citep{cooper2012}.  They matched empirical spectral templates observed with DEIMOS to the Tri~II spectra and varied the velocity until $\chi^2$ was minimized.  \citet{martin2016} used their own custom pipeline \citep{ibata2011} to reduce the spectra.  They determined radial velocities from the mean wavelengths of Gaussian fits to the Ca$\,${\sc ii} triplet.

Slit imaging spectrographs can experience radial velocity zero-point shifts if the star is not perfectly centered in the slit.  This effect can be mitigated by observing the wavelengths of telluric absorption lines \citep[e.g.,][]{sohn2007}.  The slit centering correction is taken to be the deviation of the wavelengths of these lines from the geocentric rest frame.  All of the studies used in this work performed such a correction.

\subsection{Zero-point offset}\label{sec:offset}

In this section, we describe the calculation of the zero-point velocity offset between Keck and MMT. Although our complete set of Tri~II Keck observations is a concatenation of the \cite{kirby2017} and \cite{martin2016} datasets, the zero-point offset between the two appears consistent with zero \citep[see Section 3.2 of][]{kirby2017}. For this analysis, we have defined the offset as the MMT zero-point minus the Keck zero-point. There are 8 objects in our dataset around the Tri~II galactic center with radial velocity measurements in both MMT and Keck catalogues. Measurements for a given object and instrument are combined into a single weighted mean value.  We then calculate the offset between the sets of combined instrument velocities assuming Gaussian errors. 

The methods described here and in section \ref{sec:methods} are based on a Gaussian likelihood function, which is described in terms of model prediction $\mu$, model error $\sigma$, given data in the form of the observed velocity $v$:

\begin{equation}
    \label{eq:mean_gau_like}
\mathcal{G}( v|\mu,\sigma ) =  \frac{1}{\sqrt{2\pi(v_{\rm error}^2+\sigma^2)} } \exp\left(-\frac{(v - \mu)^2}{2(v_{\rm error}^2+\sigma^2)} \right)
\end{equation}
This is the general likelihood for a given velocity prediction, and the total likelihood function is the product of the velocity likelihoods. 

In the context of finding a zero-point offset, we apply this likelihood to a set of velocity differences $\{v_i\}$, where $v_i$ represents the difference between the $i$th star's MMT weighted mean velocity and its Keck weighted mean velocity, $v_i = \bar{v}_{i, \rm MMT} - \bar{v}_{i, \rm Keck} $. We assume that these velocity differences are consistent with the zero-point offset between instruments $\delta_v$ and that there is no dispersion in $\delta_v$, only observational errors. Thus, the total likelihood is the product of the likelihoods of individual objects, $\prod_{i=1}^N \mathcal{G}(v_i, \delta_v,0)$.

However, some objects have a very large velocity difference between the instruments, possibly due to low signal-to-noise or unconfirmed binarity. A convenient way to account for these objects is to construct a mixture model to treat them as outliers, a similar model is used in \cite{li2019}. The new likelihood for a given object is the sum of the original Gaussian likelihood and outlier model likelihood weighted by the probability the object is an outlier. The new likelihood is written as... 
\begin{equation}
    \label{eq:outlier_like_model}
    \begin{split}
    L( \{v_i\}| \delta_v, \gamma, p ) =   \prod_{i=1}^N [ \gamma \mathcal{G}(v_i | 0, p ) + (1-\gamma)\mathcal{G}( v_i | \delta_v, 0 ) ]
    \end{split}
\end{equation}
where $\gamma$ represents the outlier fraction and $\delta_v$ represents the offset correction. The outlier model is applied to the difference in the combined weighted mean velocities for each instrument and is taken to be a Gaussian with a large standard deviation.  The priors are as follows:

\begin{itemize}
    \item offset $\delta_v$: uniform($-10 \rm~km~s^{-1}$, $10 \rm~km~s^{-1}$)
    \item outlier fraction $\gamma$: uniform(0, 0.5)
    \item outlier model standard deviation $p$: uniform($4\rm~km~s^{-1}$, $20 \rm~km~s^{-1}$)
\end{itemize}

We define the posterior as the Gaussian likelihood multiplied by our set of priors and look for a $\delta_v$ that maximizes this posterior. The offset error is taken as the width of this offset posterior distribution. We derive the offset value from the resulting posterior, sampled using Markov Chain Monte Carlo (MCMC) by the Metropolis–Hastings algorithm via the {\it emcee} python package \citep{foreman-mackey2013}. Applying this procedure to the 8 stars in the overlapping MMT-Keck datasets observed around the Tri~II galactic center, we find an offset value of $\delta_v = -0.11\pm 1.02 \rm~km~s^{-1}$. While this value could be determined using only our Tri~II dataset, the large error means that the offset is not resolved within $1 \rm~km~s^{-1}$ and the introduction of a large offset error can make it more difficult to resolve the Tri~II velocity dispersion.

To improve the error on our zero-point offset between the instruments, we also include Keck/DEIMOS \citep{pace2020} and MMT/Hectochelle \citep{spencer2018} observations from an additional 288 Ursa Minor objects. The combined Tri~II and Ursa Minor dataset results in an offset correction of $\delta_v = -1.33\pm 0.33 \rm km~s^{-1}$.

We opt to use the offset that is calculated while including Ursa Minor observations, as the larger amount of data gives a more precise value for the offset.  Thus, we bring all observations onto a common zero-point by subtracting a fixed amount of $1.33\pm 0.33 \rm~km~s^{-1}$ from each velocity measurement obtained with Keck/DEIMOS.  We do not propagate the offset error as the total error is dominated by the combination of instrument-specific systematics and random errors. 

\subsection{Membership}
\begin{figure}
    \centering
    \includegraphics[width=\columnwidth]{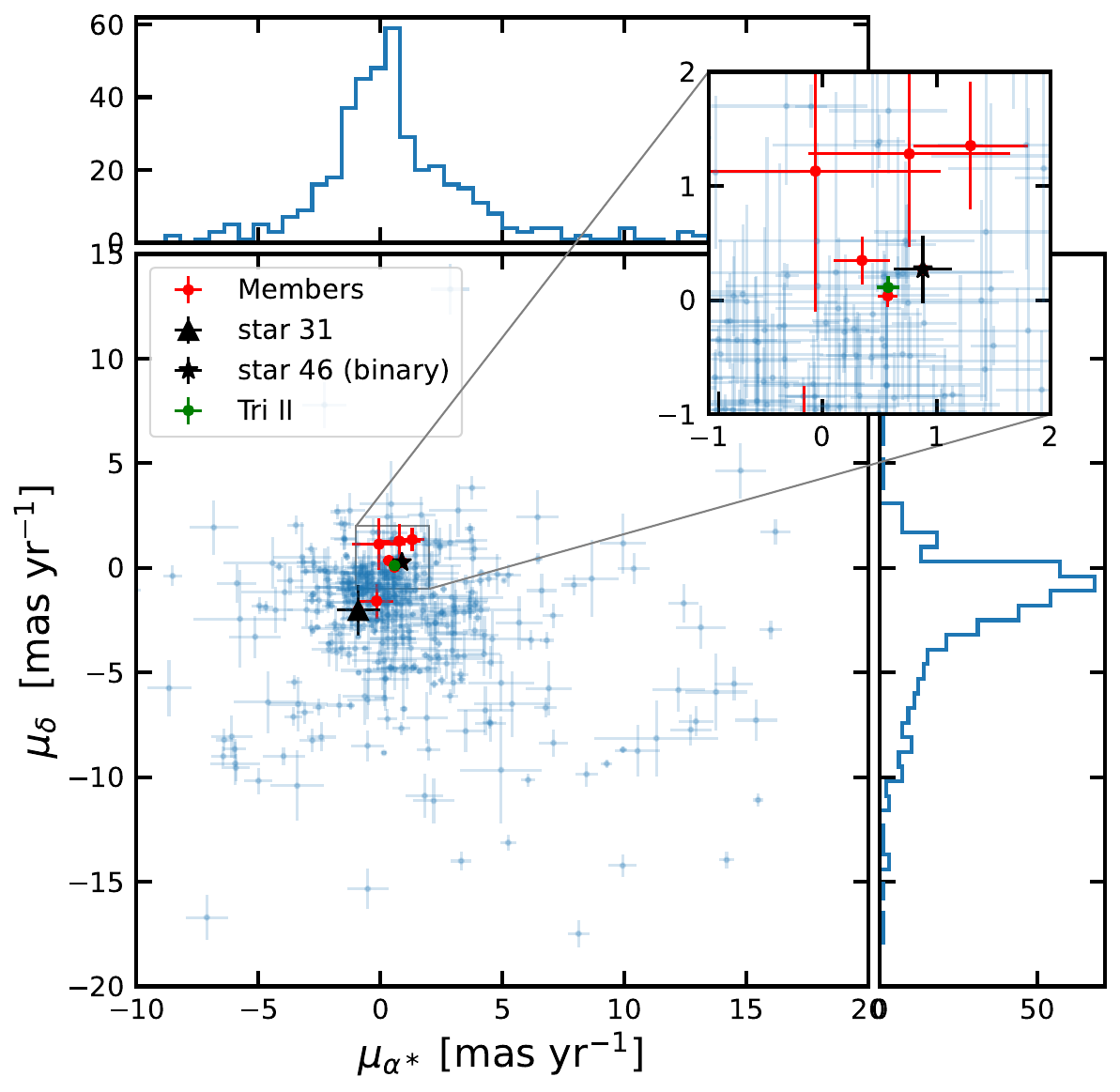}
    \caption{Proper motion distribution of stars within 10 \arcmin ($\sim 4$~ half-light radii) of Tri~II center from Gaia eDR3 after a parallax cut (blue). The green point marks the Tri~II proper motion value that is used to determine membership. The black triangle and star are the two stars with uncertain membership in \protect\cite{kirby2017}. The red dots are the remaining members that have Gaia proper motion measurements.}
    \label{fig:proper_motion}
\end{figure}

\begin{figure}
    \centering
    \includegraphics[width=\columnwidth]{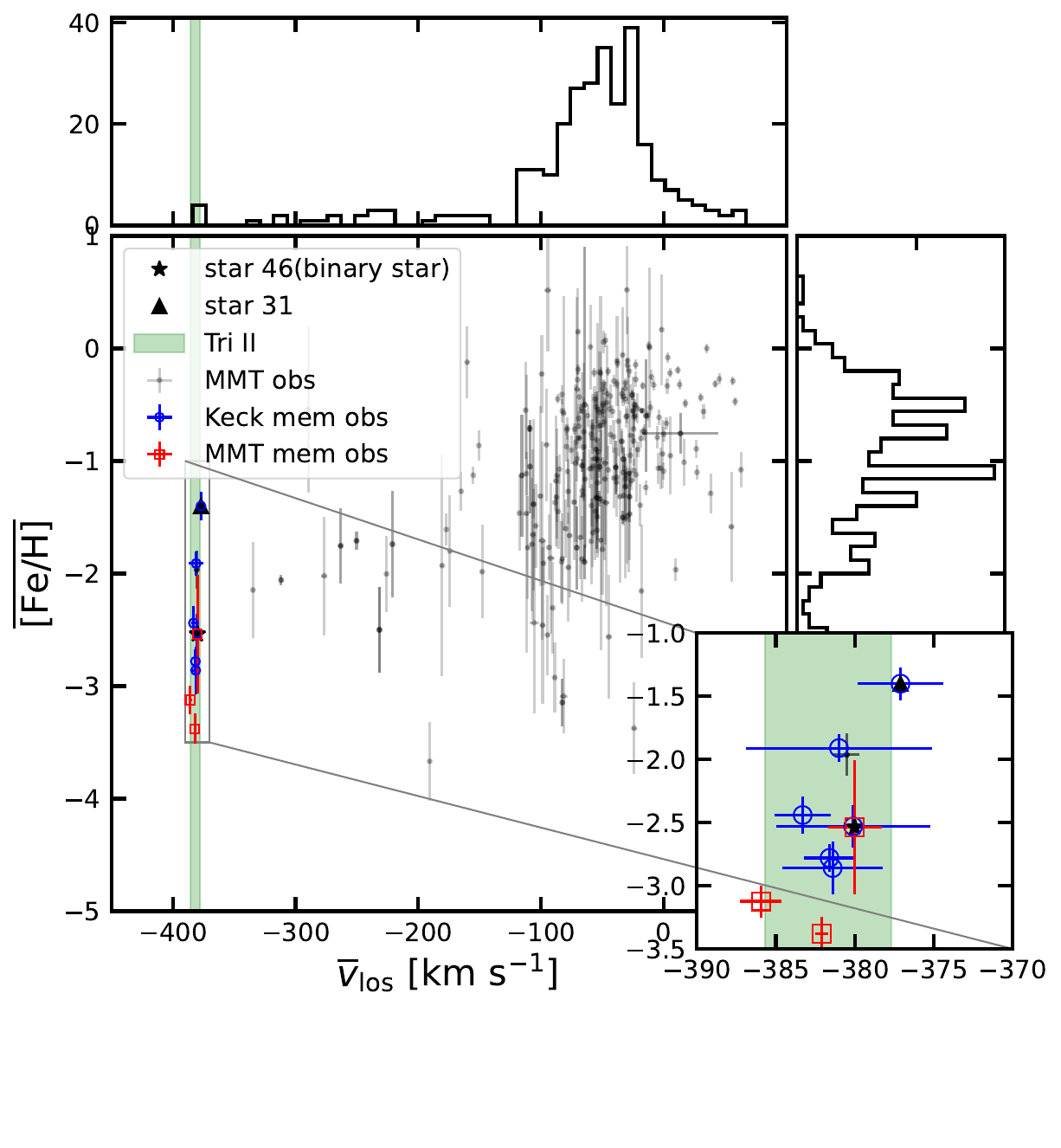}
    \caption{Mean metallicities and velocities of all stars observed in MMT around the Tri~II field (black). The green area represents the The Tri~II mean velocity from \protect\cite{kirby2017} used in membership cuts. Our final sample of Tri~II members is plotted with two colors/markers to distinguish between Keck (blue circles) and MMT (red squares) measurements. The binary star, alone, is plotted with the systemic velocity found in this work rather than the weighted mean of the velocity observations.}
    \label{fig:vel_feh}
\end{figure}

To determine Tri~II membership, we sigma clip our dataset at $3.5\sigma$ from both the galaxy mean line-of-sight velocity and proper motion.  This step removes objects whose observations differ from the Tri~II measurements $>3.5$ times the root sum of the squared of the galaxy and measurement errors, $\sigma  = \sqrt{\sigma_{\rm Tri~II}^2 + \sigma_{\rm obs}^2}$. We use the proper motions from {\it Gaia} eDR3  \citep[][Figure \ref{fig:proper_motion}]{gaiaedr3}. All members are consistent with zero parallax (i.e, $\varpi-3 \sigma_\varpi <0$). The Tri~II mean velocity is taken as $-381.7 \rm~km~s^{-1}$ \citep{kirby2017} and the galaxy proper motion as $\mu_{\alpha \star} = 0.58 \pm 0.06 \rm~mas~ yr^{-1}, \mu_{\delta} = 0.11 \pm 0.07 \rm~mas~yr^{-1}$ \citep{Pace2022arXiv220505699P}. Other systemic proper motion measurements with Gaia EDR3 find similar results \citep{mcconnachie2020updated, battaglia2022}. While the velocity dispersion is unresolved, we use $\sigma_{\rm Tri~II}=4~\rm~km~s^{-1}$ (Figure \ref{fig:vel_feh}). Our resulting sample consists of 16 member stars, one of which is the previously confirmed binary star (GAIA ID 331086526201161088 which we refer to as {\it Star46} or MIC2016-46 in this work). There is one star that is observed by MMT only, two stars in the overlap of MMT and Keck, and the remaining 13 are observed by Keck only.

\begin{figure*}
    \centering
	\includegraphics[width=\textwidth]{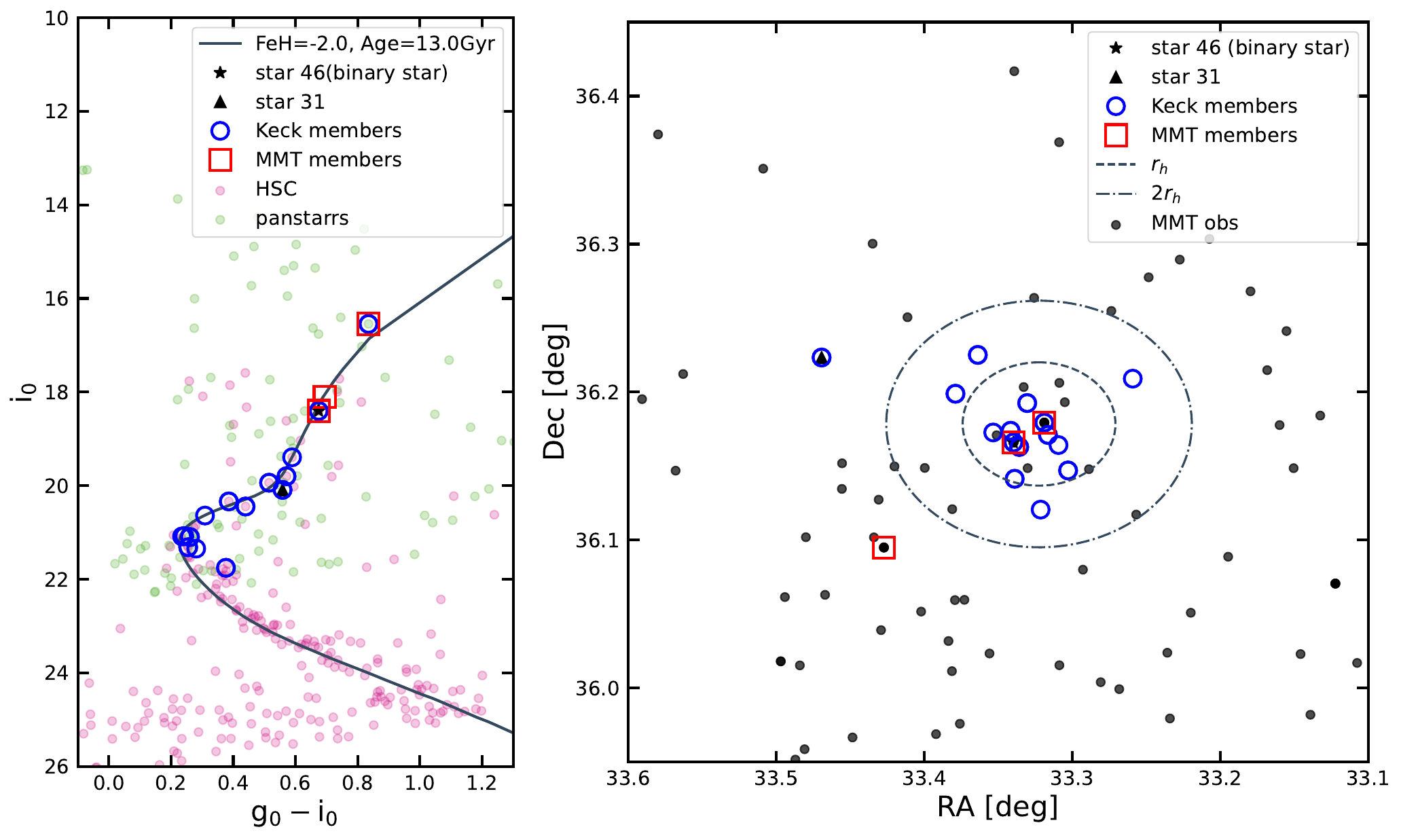}
    \caption{\textit{Left}: Color-magnitude diagram of Tri~II stars within $\sim 4$ half-light radii \citep[$r_h = $2.5\arcmin; ][]{carlin2017}  taken by Hyper Suprime-Cam (HSC) and Panstarrs with a MIST isochrone overlaid. The location of the Tri~II binary star and {\it Star31} is indicated by a star and triangle, respectively. The isochrone is defined by the following values: $m-M=17.44$, [Fe/H]=$-2.2$, and Age=13~Gyr. \textit{Right}: Sky positions of members and stars observed with MMT. The dotted lines represents the 1 and 2 half-light radii of Triangulum II. In both plots, the members are plotted with either a red square or blue circle for MMT and Keck observations respectively. }
    \label{fig:tri2_isochrone}
\end{figure*}

Figure \ref{fig:tri2_isochrone} shows the location of our final sample of member stars on a color-magnitude diagram compared to all the stars within the Tri~II half light radius after a star-galaxy separation cut observed by the Subaru/Hyper Suprime-Cam \citep[HSC;][]{HSC_cite} and Pan-STARRS1 \citep{pan-starrs1_cite}. The isochrone parameters and the half-light radius ($r_h = $2.5\arcmin) are taken from \cite{carlin2017}. In Figures \ref{fig:proper_motion}, \ref{fig:vel_feh}, and \ref{fig:tri2_isochrone}, we indicate the locations of the binary star ({\it Star46}) and another member, called {\it Star31} (or MIC2016-31) in this analysis, with black markers of different shapes.  {\it Star31} is specifically marked because of its uncertainty as a member in \cite{kirby2017} (also called {\it Star31} in their work).  Its membership is brought into question due to it being (1) far from the galactic center, (2) the most metal rich star in the sample (to the point of driving Tri~II's [Fe/H] dispersion), and (3) the only star whose velocity is $>1\sigma$ from the galaxy mean velocity. However, the now available {\it Gaia} eDR3 indicate its proper motion is consistent with Tri~II, bolstering its case as a member.

\section{Methods}\label{sec:methods}
\subsection{Binary Stars}\label{sec:binaries}
When a star is a part of a binary system, the
orbital motion can give rise to periodic variability in the line-of-sight velocity.  If one star is much brighter than its companion (e.g., a red giant with a main sequence companion), then a single-epoch spectrum may not show any evidence of variability. Conventional methods of combining observations, such as a weighted mean, can misrepresent the true systemic velocity if the system is sparsely observed and/or it results in an inflated error in the combined velocity. Thus, binary stars can directly inflate the calculated velocity dispersion if not accounted for properly. Though the true line-of-sight velocity will vary with time, it will oscillate around the center-of-mass (systemic) velocity.
 
In terms of the true anomaly $\nu$ (angular position in the orbital plane from the periastron direction), argument of periapsis $\omega$ (angle in the orbital plane between the ascending node and the periastron) and systemic velocity $v_0$, a member of a binary star system has line-of-sight velocity that varies with time according to...

\begin{equation}
\label{eq:binary_rv}
    V = K (\cos(\omega + \nu)+e \cos\omega) + v_0
\end{equation}

\noindent \citep{murrayandcorreia2010}. We can further expand the semi-amplitude $K$ in terms of the binary parameters, period $P$, semi-major axis $a$, inclination $\sin i$, and eccentricity $e$ as  

\begin{equation}
\label{eq:binary_K}
    K = \frac{2\pi}{P} \frac{a \sin i}{\sqrt{1-e^2}}.
\end{equation}

This is the model that we use to fit the observed radial velocity curves. It is worth mentioning that instead of using $\nu$ as a free parameter, we sample in $\Phi_0$, representing the phase, which is related to a constant called {\it the time of periastron passage}, $T$, by a factor of $2\pi/P$. We can relate $\nu$ and $\Phi_0$ using the {\it eccentricity anomaly}, $\zeta$.

\begin{equation}
\label{eq:anomaly}
    \tan \left( \frac{1}{2}\nu \right) = \sqrt{\frac{1+e}{1-e}} \tan \left( \frac{1}{2}\zeta \right) 
\end{equation}

\begin{equation}
\label{eq:phase0}
        \frac{2\pi}{P} (t-T) = \frac{2\pi}{P} t + \Phi_0 =  \zeta - e \sin \zeta
\end{equation}

We furthermore include a jitter parameter that acts as secondary error to account for velocity variability not due to periodic orbital motion. We note that \cite{hekker2008} found a correlation between velocity jitter and surface gravity for red-giant branch stars, where stars with $\log g \lesssim 1$ can have excess jitter $\sim 1 \rm~km~s^{-1}$. However, the majority of our members, including {\it Star46}, exist on the lower red giant branch where $\log g \gtrsim 1$. This is commonly the case for stars in ultra-faint dwarfs, so there is little concern that internal mechanisms will inflate the galaxy velocity dispersion.

We continue to assume a Gaussian error on velocity measurements, thus allowing us to use the previously defined Gaussian likelihood function, equation \ref{eq:mean_gau_like}. We generate samples of the posterior on the orbital binary parameters using rejection sampling. This sampling is carried out over the course of $10^5$ iterations, with each iteration starting with $10^6$ points before rejection. After rejection, the surviving samples across all iterations are combined. The choice to use rejection sampling instead of Monte-Carlo-Markov-Chain (MCMC) is due to the potential multi-modality posterior. A sparsely observed binary can have multiple possible orbital period solutions, which have corresponding total mass and systemic velocity values, represented as peaks in the posterior. This specific scenario can lead to untouched areas of the posterior when using MCMC.

The priors are quite uninformative, which ensures that we can fully sample all the possible binary parameters.  Note that there is a pre-defined mass range, such that the prior in total mass is loguniform from $\log(0.1 \rm~M_\odot)$ to $\log(10 \rm~M_\odot)$, which is used to bound the semi-major axis prior.  The priors on all angles are uniform. We adopt the uniform prior on $\cos(i)$ corresponding to a random orientation of binary orbits and a Beta distribution for eccentricities. The parameters for the Beta distribution are $\alpha = 1.5$ and $\beta = 1.5$, resulting in a near uniform between 0 and 1 for eccentricities, $\rm Beta(1.5,1.5)$. For compactness, we explicitly state the priors on each of the orbital parameters alongside the resulting {\it Star46} binary parameter posteriors in Table \ref{tab:binary_params}.

\begin{table*}
    \centering
    \begin{tabular}{p{0.2\linewidth}|p{0.35\linewidth}|p{0.2\linewidth}|p{0.15\linewidth}}
    \multicolumn{4}{|c|}{\textbf{Binary Star Orbital Parameters}} \\ \hline \hline
    Parameter & Prior & Short Per. & Long Per.\\ \hline
Period $P$ & linearly decreasing in log($P$) from 1 day to $10^{10}$ days & $148.1_{-1.4}^{+2.8}$ days & $296.1_{-3.3}^{+3.8}$ days\\
&&&\\
Eccentricity $e$ & Beta(1.5,1.5) & $0.29_{-0.07}^{+0.10}$  & $0.50_{-0.04}^{+0.06}$ \\
&&&\\
Systemic velocity $v_0$ & Normal(0,$474\rm~km~s^{-1}$) & $-387.6_{-1.2}^{+0.8}$ $\rm~km~s^{-1}$ & $-380.0_{-1.7}^{+1.8}$ $\rm~km~s^{-1}$\\
&&&\\
Phase $\Phi_0$ & uniform(0,2$\pi$) & $3.12_{-2.10}^{+2.21}$  & $3.25_{-2.23}^{+2.04}$ \\
&&&\\
Argument of periapsis $\omega$ & uniform(0,2$\pi$) & $4.04_{-0.65}^{+0.35}$  & $3.43_{-0.24}^{+0.28}$ \\
&&&\\
Jitter & Normal($0\rm~k~ms^{-1}, 2\rm~k~ms^{-1})$ & $1.3_{-0.9}^{+1.4}$ $\rm~km~s^{-1}$ & $0.9_{-0.6}^{+1.1}$ $\rm~km~s^{-1}$\\
&&&\\
Semi-major axis $a$ & loguniform, bounds determined from mass & $0.71_{-0.15}^{+0.26}$ AU & $1.12_{-0.24}^{+0.41}$ AU\\
&&&\\
Inclination $\sin i$ & uniform(0,2$\pi$) in $\cos i$ & $0.31_{-0.09}^{+0.08}$  & $0.39_{-0.11}^{+0.11}$ \\
&&&\\
    \hline
    \end{tabular}
    \caption{Priors used in binary modeling and resulting orbital parameters for the Tri~II binary star. The posterior is separated into short and long period solutions at the 0.6 years (219.2 days) boundary.}
    \label{tab:binary_params}
\end{table*}

We are able to make further assumptions about the system by modifying the surviving posterior samples. We remove nonphysical solutions corresponding to close binaries where the pericenter is less than the stellar radius or where the mass of the companion star is negative. These stellar radii and masses are calculated using the {\it Isochrones} python package \citep{morton2015} to determine stellar parameters for the binary star of age 13 Gyr, [Fe/H] of -2.2, and distance modulus of 17.24 using the MESA Isochrones and Stellar Tracks (MIST) grid  (\citealt{MISTisochrone0}; \citealt{MISTisochrone1}).  Figure \ref{fig:tri2_isochrone} shows the relevant isochrone fit on a color-magnitude diagram with Tri~II members plotted. For the binary star {\it Star46}, we determine a mass of $M_{\rm bin}=0.777 \rm ~M_{\odot}$ and stellar radius of $r_{\rm bin}= 0.027 \rm~AU$ from an isochrone fit to Tri~II stars.  We equivalently derive a mass and radius for the other Tri~II members by finding the point on the isochrone closest to each member star using the same steps. 

Similarly, \cite{raghavan2010} looked at 259 confirmed solar-type binaries in the \textit{Hipparcos} catalog and found that the periods are distributed according to a log-normal distribution with $\mu_{\log P} = 5.03$ and standard deviation $\sigma_{\log P} = 2.28$, where $P$ is in days. This is a more informative prior on the period than our default period prior. We refer to this lognormal distribution as the "Raghavan prior" for the remainder of this paper. We present this distribution with the minor caveat that most of the detectable member stars of dwarf galaxies are not solar-type (main-sequence) stars, but rather red giant branch stars. 

We are able to infer the posterior under this prior by re-sampling from the current surviving samples with non-uniform weights. These weights are proportional to the ratio of the Raghavan period distribution to the default prior. All of these assumptions shift the posterior away from shorter orbital periods, but as we will see in section \ref{sec: galaxy_dynamics}, the application of this prior on {\it Star46} significantly alters the parameter posterior. As such, we present the results both before and after the application of the Raghavan prior.

\subsection{Non-Binary Stars}\label{sec:nonbinaries}
\label{sec:non-var_stars}
In the case of non-binary stars, we expect line-of-sight velocity observations to be consistent with a constant velocity model with some scatter in each measurement proportional to the observational error. Our goal is to determine a star's true velocity by combining multiple epochs of radial velocity measurements into one stellar velocity. A common approach to this is calculating the weighted mean---i.e., summing the measurements weighted by the inverse of the squared errors. This approach has the benefits of being analytic and thus not computationally intensive. However, it has the downside of underestimating the error when dealing with potentially variable systems. In this section, we offer another method to calculate the true stellar velocity by finding its posterior distribution. 

\begin{figure}
	\includegraphics[width=\columnwidth]{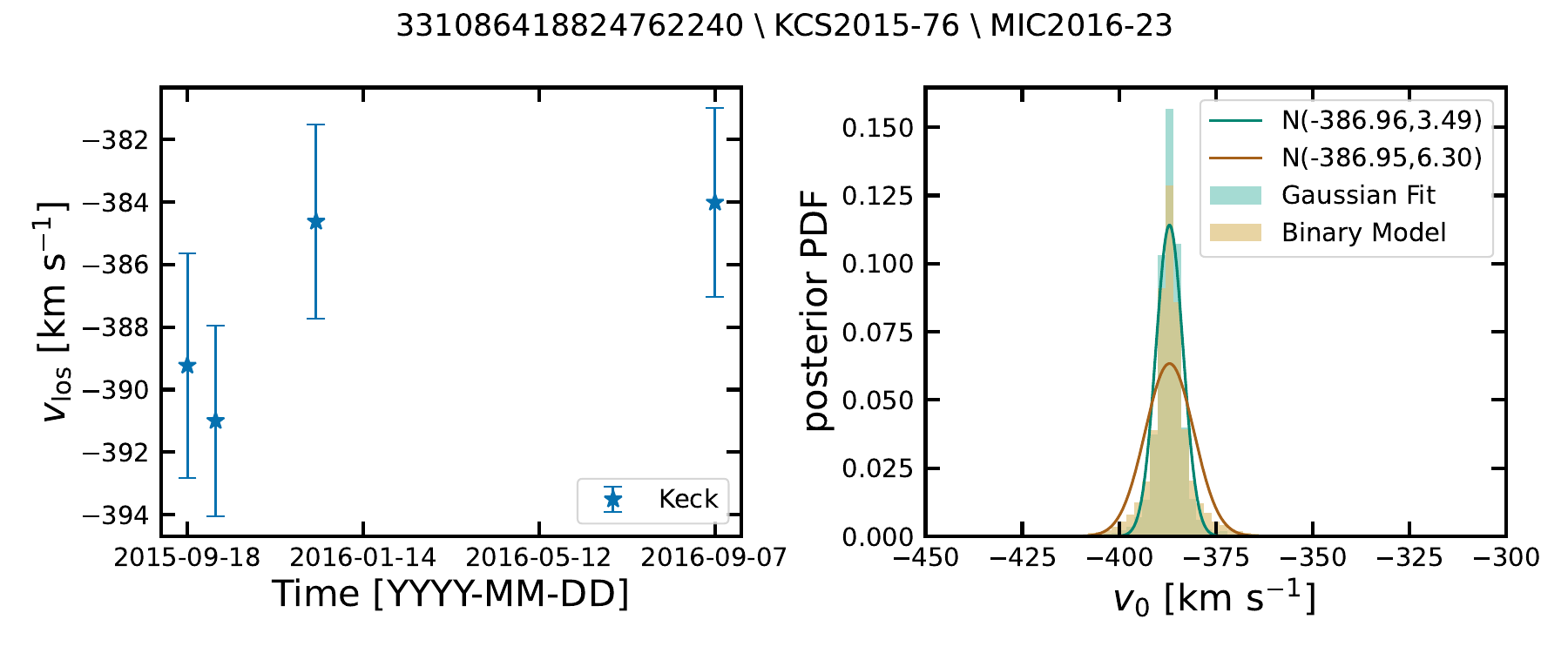}
    \caption{\textit{Left:} RV curve of \textit{Star 23}, a Tri~II member with more than one observation. The velocity measurements are colored by instrument with the same labeling as Figure \ref{fig:nonbinary_rvs}. \textit{Right:} Corresponding posterior on the mean velocity resulting from both the Gaussian fit and binary model. The binary model posterior samples have had both modifications that remove short period solutions to reduce the posterior scatter. }
    \label{fig:nonbinary_post}
\end{figure}

Once again, the assumption of Gaussian error allows us to use the previously defined Gaussian likelihood function, equation \ref{eq:mean_gau_like}, and a set of priors in true velocity and velocity error to sample from the true velocity posterior. The sampling is done using MCMC sampling and is performed with 24 walkers, 1000 steps each, and 100 step burn-in. For a set of line-of-sight velocity observations $\{v_i\}$ with errors $\{\epsilon_i\}$, we define a uniform prior in true velocity that is non-zero from $\min(\{v_i\}) - 4 \max(\{\epsilon_i\}) $ to $\max(\{v_i\}) + 4 \max(\{\epsilon_i\}) $. We also use Jeffrey's prior for the standard deviation of a Gaussian distribution ($1/\sigma$ or uniform in $\log\sigma$) for the true velocity error. The error of the combined velocity measurement is taken as one standard deviation from the posterior on the mean. We use the mixture sub package found in {\it scikit-learn} \citep{scikit-learn} to fit a Gaussian mixture model (GMM) of a singular Gaussian to the resulting posterior samples. Figure \ref{fig:nonbinary_post} shows the radial velocity curve and resulting posteriors using this method, referred to as the Gaussian fit method, and binary model method for an example Tri~II member star with multiple velocity observations.

\begin{figure*}
\includegraphics[width=\textwidth]{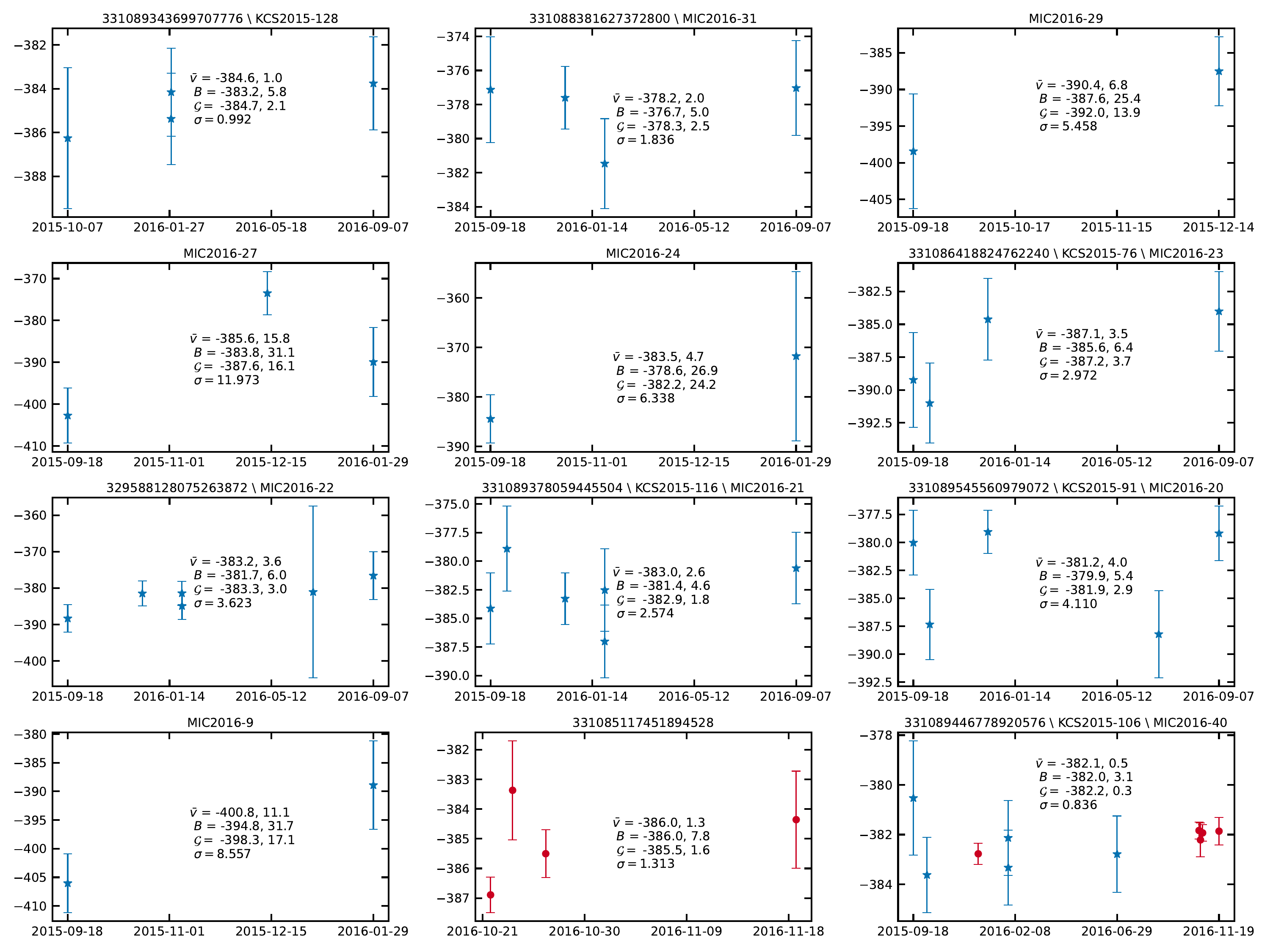}
\caption{Radial velocity curves of the 12 non-binary member stars with multiple epochs. The different values are the results of different methods of combining observations. $\bar{v}$: weighted mean; $B$: fitting a Gaussian distribution to the binary model $v_0$ posterior; $\mathcal{G}$: assuming the observations are taken from a Gaussian distribution centered on a true systemic velocity. For comparison, the standard deviation of the observations is also listed ($\sigma$). The stars are identified by their Gaia ID when applicable. All values listed and on the y-axis are in $\rm km~s^{-1}$. The dates are in the format Year-Month-Day.}
    \label{fig:nonbinary_rvs}
\end{figure*}

This Gaussian Fit method has the advantage in that it does not assume binarity but does retain increased error with larger variability in the RV curve, more than the weighted mean. For comparison, we also calculate the combined velocities under the weighted mean and binary model methods (Figure \ref{fig:nonbinary_rvs}). Applying the binary model to the, assumed non-binary, member stars takes up significantly more computation time and results in a larger uncertainty in the combined velocity. Over short time-scales, the observations can be consistent with either a constant velocity model or the RV curve of a long period binary. These methods are only applied to members with more than one velocity measurement as neither would be informative for stars with only a single epoch. Using the Gaussian Fit method would be the equivalent of trying to fit a Gaussian distribution to a single point. For the binary model, there is not enough information in a single epoch to restrict the possible binary solutions.

\subsection{Mean Velocity and Velocity Dispersion} \label{sec:velocity dispersion}
\begin{figure}
	\includegraphics[width=\columnwidth]{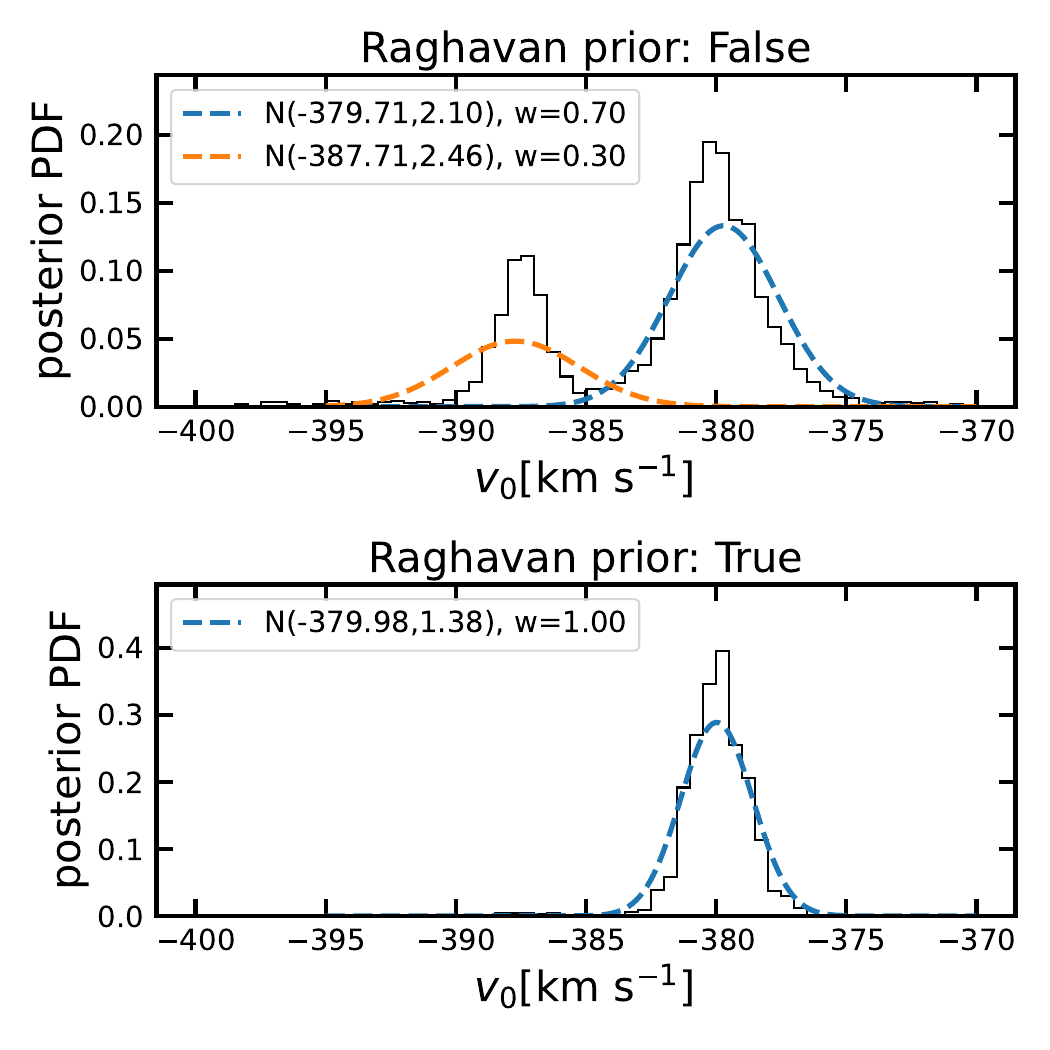}
    \caption{The posterior on the systemic velocity of the binary star and the Gaussian mixture model fit under different conditions. These conditions are a re-weight of the samples such that the period prior is the Raghavan distribution and removing non-physical samples.}
    \label{fig:binary_v0}
\end{figure}

The Tri~II velocity dispersion can be inferred from the posterior using the same Gaussian likelihood function as in previous sections (Equation \ref{eq:mean_gau_like}). We define another Gaussian model in terms of the mean velocity $\bar{v}$ and the velocity dispersion $\sigma_v$ of the galaxy. The total likelihood is the product of the likelihood of the model on the stellar velocity of each member. For stars with multiple velocity measurements, the measured velocities are combined using one of the methods presented in the previous sections. In the case of a multi-modal posterior, such as for {\it Star46}, we must fit a one-dimensional GMM to the systemic velocity posterior and the likelihood for that object becomes...
\begin{equation}
    \label{eq:gmm_like}
    \centering
    L =   \sum_{j}^n  w_j \mathcal{G}( \mu_j | \bar{v}, \sigma_v )
\end{equation}
where $\mu_j$ and $w_j$ represent the mean and weight of the $j$th Gaussian in the mixture model with $n$ Gaussians. The error on $\mu_j$ is taken as the standard deviation of the Gaussian $\sigma_j$ and the number of Gaussians in the GMM is determined by manually inspecting the $v_0$ posterior. Figure \ref{fig:binary_v0} shows the GMM fits to the {\it Star46} systemic velocity posterior before and after the application of the Raghavan prior. 

The velocity dispersion prior is log-uniform from $\log_{10}(0.05 \rm~km~s^{-1})$ to $\log_{10}(100 \rm~km~s^{-1})$, which is equivalent to Jeffrey's prior. The minimum value corresponds to the case where the galactic dynamics are fully determined by visible matter (see Appendix \ref{appendix: sigv_prior}). We have chosen to find the posterior of $\log_{10}\sigma_v$ rather than $\sigma_v$ to better sample smaller velocity dispersion values. Once again, we use a uniform prior from $\min(\{v_i\}) - 4 \max(\{\epsilon_i\}) $ to $\max(\{v_i\}) + 4 \max(\{\epsilon_i\}) $ as our prior on mean velocity.

\section{Results}\label{sec:results}
\subsection{Binary Orbital Parameters}\label{sec:orbital_params}

Within our data set, we are able to spectroscopically identify one star, {\it Star46}, as a binary. Attempting to fit a constant velocity model to the observations of this object, we find a reduced chi-squared of $ \chi_{\rm red} = 157.2$.  Previous analyses by \cite{venn2017} and \cite{kirby2017} found this star to be part of a binary system. For comparison, the member with the second highest reduced chi-squared is MIC2016-27 with $ \chi_{\rm red} = 4.4$. Although this value could suggest possible binarity for MIC2016-27, its velocity variations are more consistent with a constant-velocity model than {\it Star46}.

In this analysis, we use the 4 velocity measurements of this star from MMT (Table \ref{tab:obs_MMT}) and 5 from Keck (Table \ref{tab:obs_keck}). There are two additional velocity measurements made in Geminin/GRACES data \citep{venn2017, ji2020}  that we have opted to not use for two reasons. First, the inclusion of this data would add an additional unknown zero-point correction that is more difficult to quantify due to less overlap in number of stars. Second, we have explored including them and found that they do not impact the resulting orbital parameter posterior (See Appendix \ref{appendix: GRACES}).

\begin{figure}
    \includegraphics[width=\columnwidth]{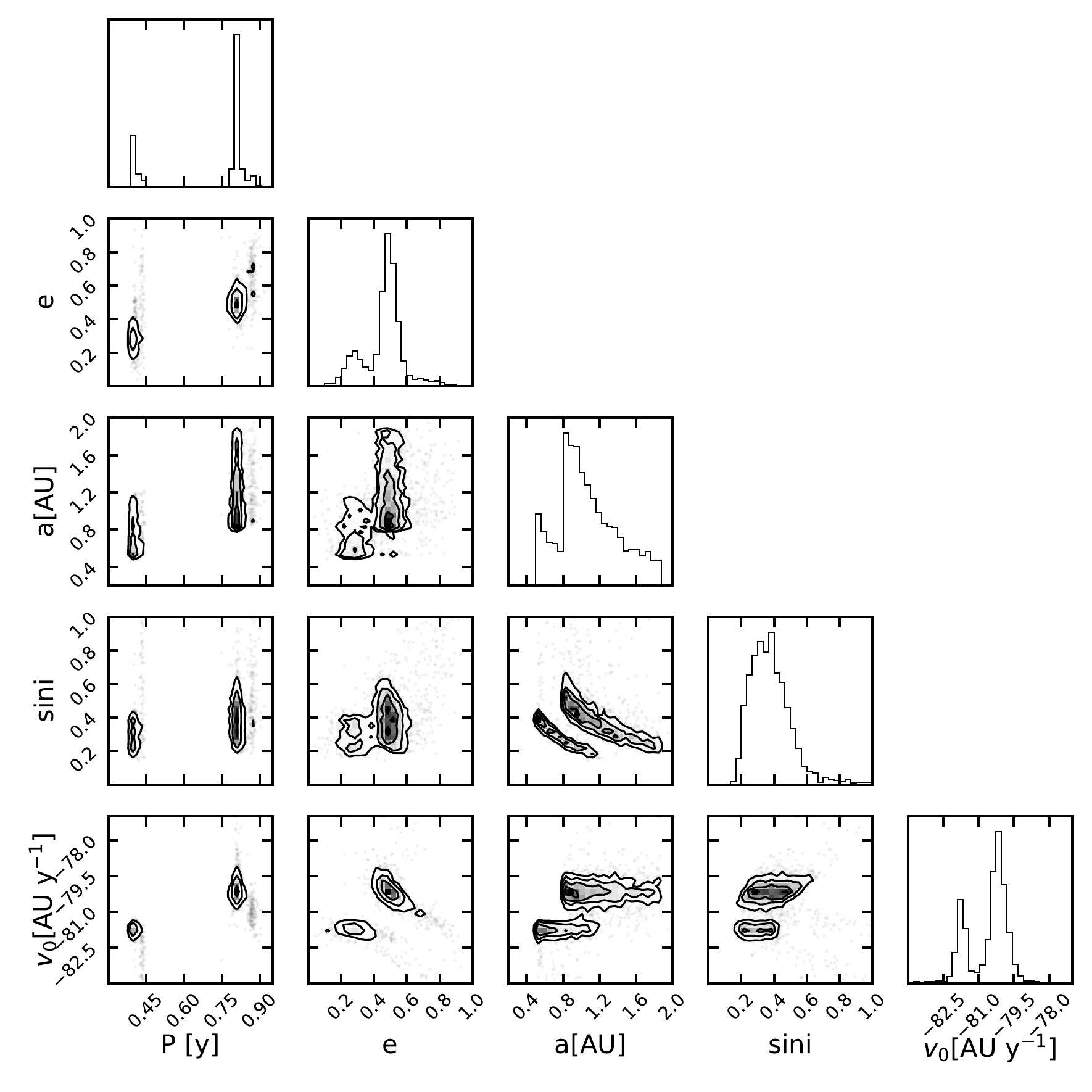}
    \caption{Orbital parameters' posterior for the binary star derived from our binary model. There are two distinct modes corresponding to }
    \label{fig:binary_corner}
\end{figure}

\begin{figure}
	\includegraphics[width=\columnwidth]{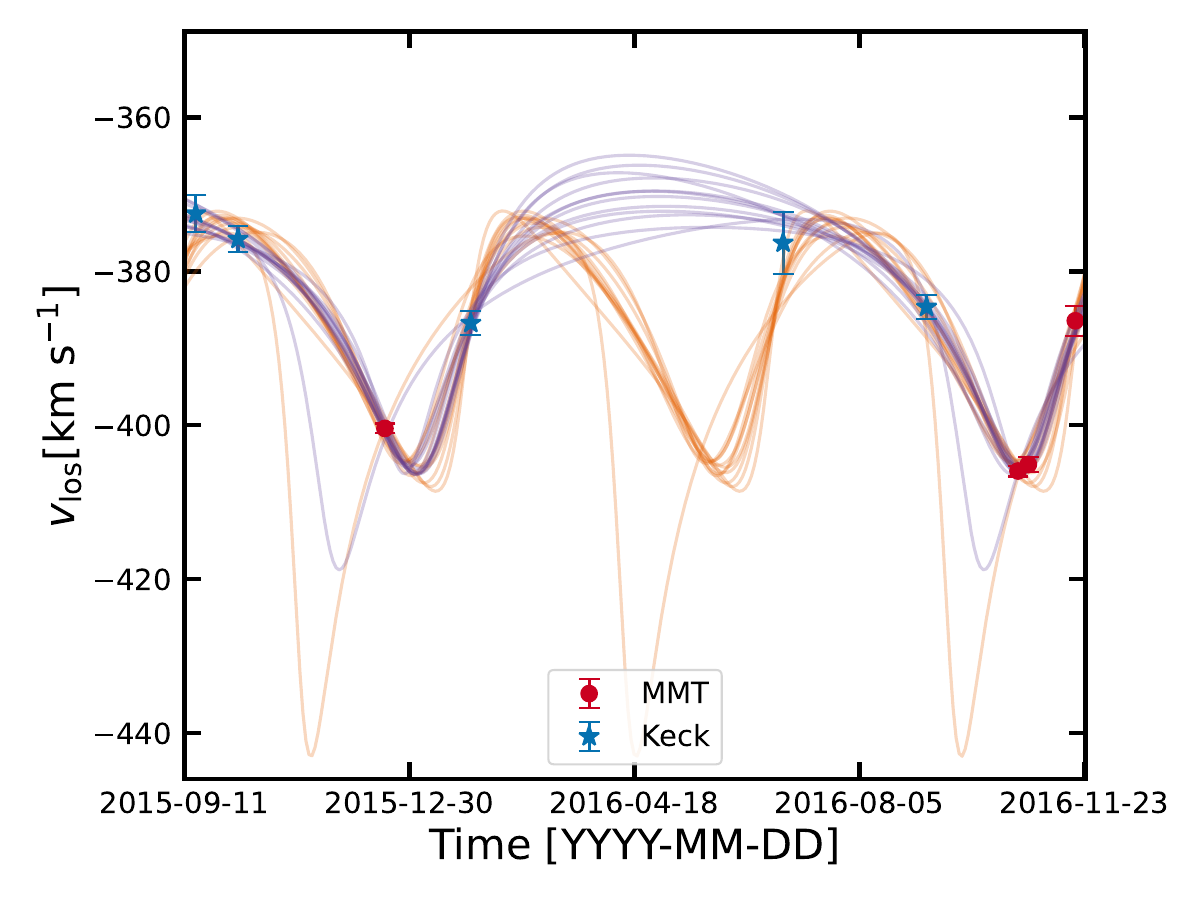}
    \caption{Possible fits to the radial velocity curve of the binary star, {\it Star46}, calculated using our binary model. The orange and purple lines correspond to solutions with orbital periods of approximately 0.4 years and 0.8 years respectively. The dates are in the format Year-Month-Day.}
    \label{fig:binary_rv}
\end{figure}

After an initial pass with our model on {\it Star46}, we find that there are two noticeable peaks in the posterior corresponding to periods of $\sim$ 0.4 and $\sim$ 0.8 years. We are able to eliminate the sub-harmonics (e.g., 0.2, 0.1 years) from our restrictions on pericenter and Kepler's third law both limiting the allowed semi-major axis values. Because the period modes are distinct, we re-run the sampling with a more restricted period prior bound at 0.3 and 0.95 years to ensure a larger number of surviving posterior samples. The resulting binary orbital parameters presented in this section are derived from the posterior after removing samples that correspond to a negative companion mass but before the application of the Raghavan prior. Figure \ref{fig:binary_corner} is a corner plot of the posterior of select parameters and shows two distinct peaks corresponding to long and short period solutions. We separate the posterior samples at the 0.6 year boundary and present the orbital parameter values taken from the truncated posteriors in Table \ref{tab:binary_params}. Figure \ref{fig:binary_rv} shows the RV curve for this star, as well as the orbital solutions that correspond to the surviving samples. 

The posteriors of the semi-major axis for both solutions are consistent with values less than $a = 2 \rm AU$, indicating that this is a close binary. The resulting overall binary parameter space is that of a common binary star system. The companion mass posterior is not distinct enough from the prior to be informative, but all of the samples correspond to stellar masses potentially on the main sequence.

For comparison, the only other binary star in an ultra-faint dwarf spheroidal, Her-3 in the Hercules  dwarf, has a smaller eccentricity ($e_{\rm Her-3} = 0.18$) and period ($P_{\rm Her-3} = 135.28 \pm 0.33 \rm days$) values \citep{koch2014}. Plugging into equation \ref{eq:binary_K} we find the corresponding semi-amplitudes, $K_{\rm short} = 16.24_{-1.17}^{+1.59}~\rm km~s^{-1}$ and $K_{\rm long} = 18.72_{-1.59}^{+2.67}~\rm km~s^{-1}$ which are only slightly larger than Herc-3's solution, $K_{\rm Her-3} = 14.48 \pm 0.82 ~\rm km~s^{-1}$. 

The truncation in companion mass slightly shifts the posterior towards longer periods, but not more than $0.2 \rm~km~s^{-1}$, much smaller than the width of the posterior peaks ($\sim 2 \rm~km~s^{-1}$). Reweighting the samples so that the prior is the Raghavan period distribution weights the posterior in favor of the longer period solutions and the corresponding larger systemic velocities, removing the multi-modality of the posterior (Figure \ref{fig:binary_v0}). Even before applying the Raghavan prior, there is a strong preference towards the long period solutions as the ratio of the long period posterior to short period posterior is approximately 5:2. After applying the Raghavan prior, the posterior becomes almost entirely in favor of the long period solution as the ratio becomes 80:1, effectively removing the short period mode. As such, we present the parameters corresponding to the long period mode as the orbital solution, but we also explore the effect using the Raghavan prior has on the velocity dispersion.

This binary star is only 1 of the 16 stars in our sample. \cite{moe2019} and \cite{mazzola2020} found that the intrinsic close binary fraction is anticorrelated with metallicity such that Tri~II's metallicity value of [Fe/H]=-2.1 implies a close binary fraction $f_{\rm bin, close} \approx 0.5$. This suggests that there are more close binary members, but the current data is not enough to determine binarity for any other binaries in the galaxy. Though, with growing observation power, they may one day be observed sufficiently.

\subsection{Triangulum II Dynamics}\label{sec: galaxy_dynamics}

While we have briefly explored applying the binary model to the non-binary members and applying the Gaussian fit model to {\it Star46}, we unsurprisingly find that the best variation is to treat confirmed the binary star under the binary model and the remaining members under the Gaussian fit model. This option assumes variability for only {\it Star46}. These additional variations made it more difficult to resolve the dispersion and instead resulted in a posterior with a maximum at the $\sigma_v$ prior's minimum value. Thus, we focus our final results on the best variation.

The galaxy mean velocity posterior is consistent with the previously found mean velocity of Tri~II, $\langle v \rangle  = -381.7 \pm 1.1 \rm~km~s^{-1}$ \citep{kirby2017} within $2 \sigma$, where $\sigma$ is the error on the mean velocity from the posterior. We present a mean velocity of $\langle v \rangle  = -382.3 \pm 0.7 \rm~km~s^{-1}$ from this analysis.  However, even in this best variation, we were unable to completely resolve the velocity dispersion.

\begin{figure}
	\includegraphics[width=\columnwidth]{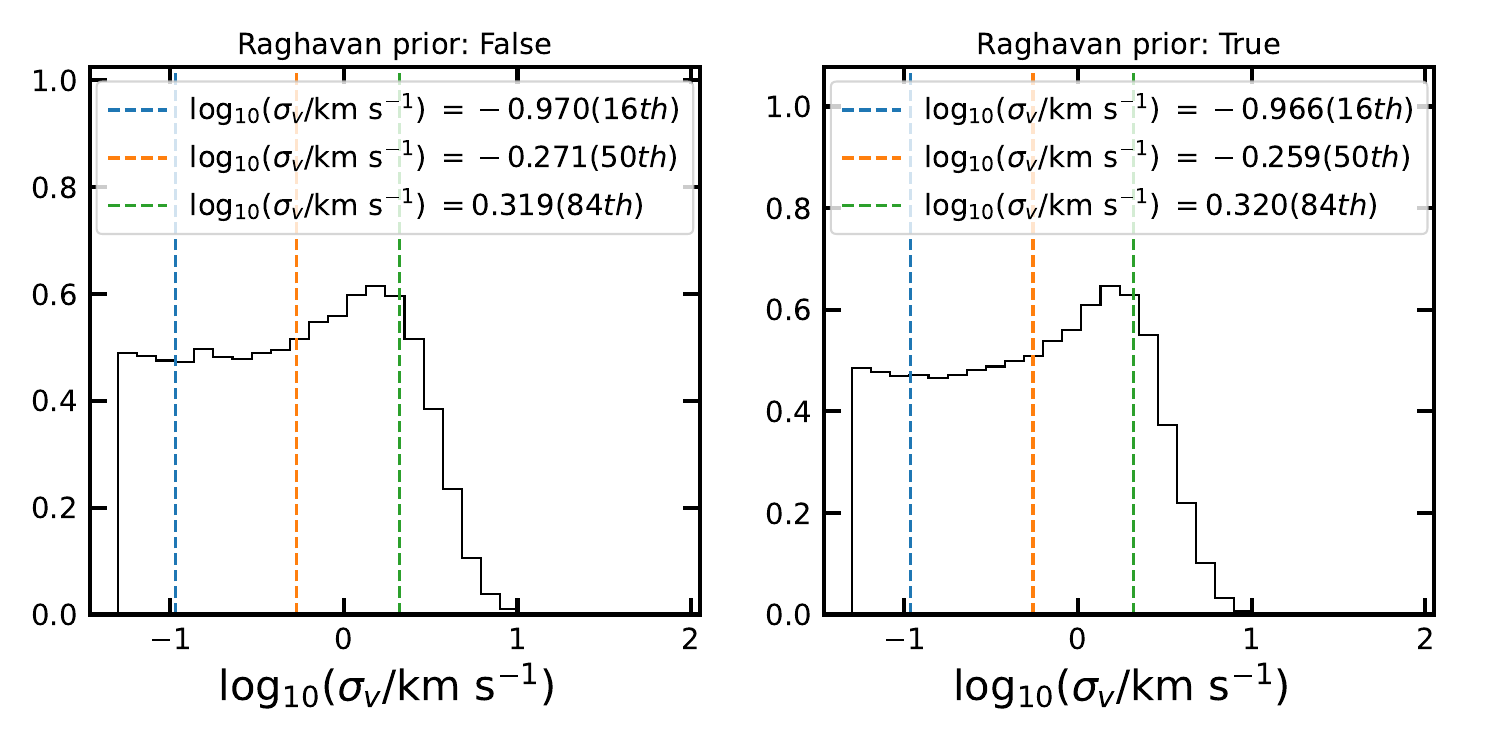}
    \caption{Velocity dispersion posterior with and without the use of the Raghavan prior on {\it Star46}. The 16th, 50th, and 84th percentiles are marked by the blue, orange and green lines respectively.}
    \label{fig:vel_disp_condtional}
\end{figure}

\begin{figure}
	\includegraphics[width=\columnwidth]{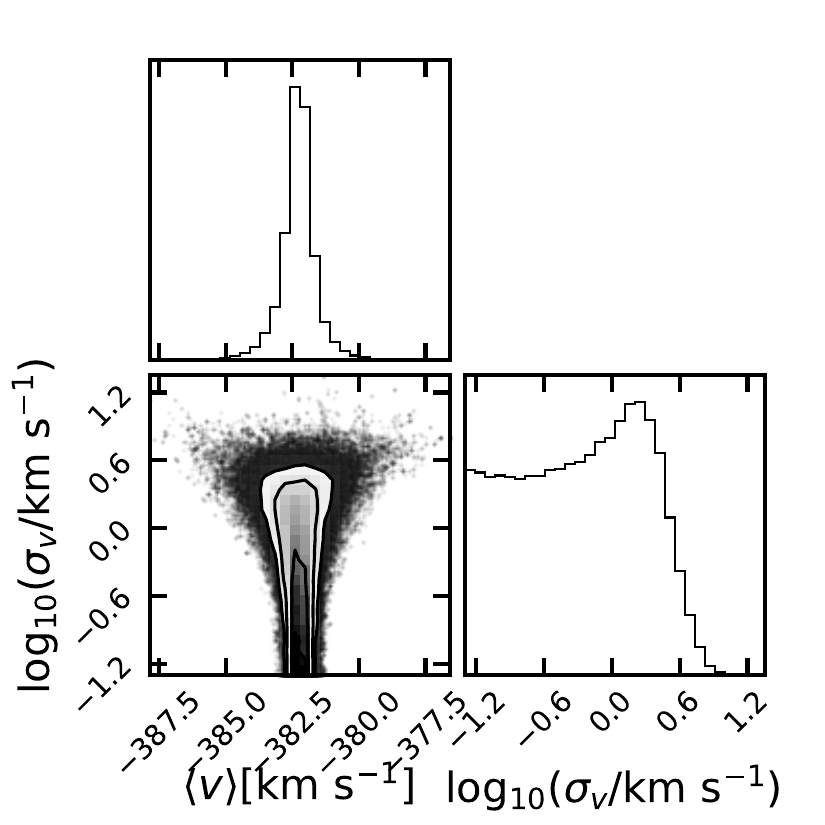}
    \caption{Complete posterior Tri~II mean velocity and velocity dispersion when treating the binary star under the binary model and other non-binary members under the Gaussian fit with the Raghavan prior.}
    \label{fig:tri2_posterior}
\end{figure}

The posterior has a small peak at $\log_{10}\sigma_v = 0.2$ (which corresponds to  $\sigma_v = 1.6 \rm km~s^{-1}$), but a  large probability tail that stretches back to the minimum value. We see that applying the Raghavan prior only slightly reduces this tail (Figure \ref{fig:vel_disp_condtional}).  Figure \ref{fig:tri2_posterior} shows the complete posterior with the binary under the Raghavan prior. This remaining tail of probability means we were unable to resolve the velocity dispersion of Tri~II, and we instead present upper limits on $\sigma_v$. Before applying the Raghavan prior, we find a 90\% confidence limit of $2.7 \rm~km~s^{-1}$ and 95\% confidence limit of $3.5 \rm~km~s^{-1}$. After applying the Raghavan prior, we find a 90\% confidence limit of $2.6 \rm~km~s^{-1}$ and 95\% confidence limit of $3.4 \rm~km~s^{-1}$. Due to the differences between these values being so small, we again choose to present the values determined using the more informative period prior.

\section{Conclusions}\label{sec:conclusions}

We have found orbital parameters for the Tri~II binary star member, {\it Star46}. This is now the second binary star within an ultra-faint dwarf with an orbital solution after Her-3 in Hercules Dwarf \citep{koch2014}. In doing so, we have demonstrated a method of accounting for binary star systems in velocity dispersion calculations that does not demand the removal of such systems. We also see that using our binary model when unnecessary, such as with non-binary stars, not only adds lengthy computation time, but can lead to an overestimation of velocity error.

The classification of Tri~II as either an ultra-faint dwarf or globular cluster remains an open question. Though the use of this model has offered an improved analysis of the dwarf galaxy, we were unable to resolve a velocity dispersion that would confirm the existence of a dark matter halo in Tri~II. However, \cite{kirby2017} found a metallicity dispersion of $\sigma([\rm Fe/H]) = 0.53^{+0.38}_{-0.12} \rm dex$ when including all available metallicity measurements for the dataset. They make clear that metallicity dispersion relied on the membership two stars, {\it Star31} and {\it Star46}, despite knowing that {\it Star46} is in a binary system. The {\it Gaia} proper motions of both stars confirm that they are Tri~II members  \citep[Figure \ref{fig:proper_motion}; also see][and Pace et al. 2022 in prep.]{mcconnachie2020updated}. We briefly tested performing the same $\sigma([\rm Fe/H])$ calculation while including the new MMT metallicities. We found that the inclusion of MMT measurements did not significantly impact the resulting dispersion where the difference after inclusion was $\lesssim 0.03 \rm ~dex$ from the previous value (See Appendix \ref{appendix: feh_disp}). This metallicity dispersion greatly strengthens the case for Tri~II's classification as a dwarf galaxy, indicating the system is embedded in a dark matter halo \citep{willman_strader2012}.

The choice to apply our binary model to only one star means that we are operating with a hard binary fraction of $f_{\rm bin} \approx 0.06$. A more informed model would be to assume that there is some non-zero fraction of the ``non-binary'' member stars that are actually binaries. Each star would have an associated binary and non-binary likelihoods that are calculated from marginalizing over the binary parameter posterior. These likelihoods would allow for the sampling of binary fraction as a parameter alongside with mean velocity and velocity dispersion, forming a hierarchical model that builds from the set of binary posteriors. This likelihood would be functionally similar to the mixture model used to determine the offset between the instrumental zero-point velocities. However, this approach introduces complications, as marginalization becomes less straightforward when modifying the posterior. For instance, removing binary samples based on a parameter, such as pericenter, makes the posterior no longer normalized to the calculated binary/non-binary likelihoods. The likelihoods must be re-normalized after truncation for every additional condition that is imposed. Though it is outside the scope of this paper, we find the hierarchical model to be an interesting problem and important step to explored in future works.

\section*{Acknowledgements}
This work is supported in part by NSF grants AST-1813881 and AST-1909584.  E.N.K.\ is supported by NSF grant AST-1847909, and he gratefully
acknowledges support from a Cottrell Scholar award administered by the
Research Corporation for Science Advancement.
EO is partially supported by NSF grant AST-1815767.
NC is supported by NSF grant AST-1812461.
MM is supported by U.S.\ National Science Foundation (NSF) grants AST-1312997, AST-1726457 and AST-1815403.  
ES acknowledges funding through VIDI grant "Pushing Galactic Archaeology to its limits" (with project number VI.Vidi.193.093) which is funded by the Dutch Research Council (NWO).
CB and CMD are supported by NSF grant AST-1909022.

This work has made use of data from the European Space Agency (ESA)
mission {\it Gaia} (\url{https://www.cosmos.esa.int/gaia}), processed by the {\it Gaia} Data Processing and Analysis Consortium (DPAC,
\url{https://www.cosmos.esa.int/web/gaia/dpac/consortium}). Funding
for the DPAC has been provided by national institutions, in particular
the institutions participating in the {\it Gaia} Multilateral Agreement.

The Pan-STARRS1 Surveys (PS1) and the PS1 public science archive have been made possible through contributions by the Institute for Astronomy, the University of Hawaii, the Pan-STARRS Project Office, the Max-Planck Society and its participating institutes, the Max Planck Institute for Astronomy, Heidelberg and the Max Planck Institute for Extraterrestrial Physics, Garching, The Johns Hopkins University, Durham University, the University of Edinburgh, the Queen's University Belfast, the Harvard-Smithsonian Center for Astrophysics, the Las Cumbres Observatory Global Telescope Network Incorporated, the National Central University of Taiwan, the Space Telescope Science Institute, the National Aeronautics and Space Administration under Grant No. NNX08AR22G issued through the Planetary Science Division of the NASA Science Mission Directorate, the National Science Foundation Grant No. AST-1238877, the University of Maryland, Eotvos Lorand University (ELTE), the Los Alamos National Laboratory, and the Gordon and Betty Moore Foundation.

This research has made use of NASA's Astrophysics Data System Bibliographic Services.
This paper made use of the Whole Sky Database (wsdb) created by Sergey Koposov and maintained at the Institute of Astronomy, Cambridge by Sergey Koposov, Vasily Belokurov and Wyn Evans with financial support from the Science \& Technology Facilities Council (STFC) and the European Research Council (ERC).

Observations reported here were obtained at the MMT Observatory, a joint facility of the University of Arizona and the Smithsonian Institution. The authors wish to recognize and acknowledge the very significant cultural role and reverence that the summit of Maunakea has always had within the indigenous Hawaiian community.  We are most fortunate to have the opportunity to conduct observations from this mountain.

For the purpose of open access, the author has applied a Creative Commons
Attribution (CC BY) licence to any Author Accepted Manuscript version arising
from this submission.
\section*{Data Availability}

The processed MMT/Hectochelle spectra for Triangulum II targets are publicly available at the Zenodo database, \url{https://doi.org/10.5281/zenodo.6561483}.



\bibliographystyle{mnras}
\bibliography{references} 

\begin{thebibliography}{}
\makeatletter
\relax
\def\mn@urlcharsother{\let\do\@makeother \do\$\do\&\do\#\do\^\do\_\do\%\do\~}
\def\mn@doi{\begingroup\mn@urlcharsother \@ifnextchar [ {\mn@doi@}
  {\mn@doi@[]}}
\def\mn@doi@[#1]#2{\def\@tempa{#1}\ifx\@tempa\@empty \href
  {http://dx.doi.org/#2} {doi:#2}\else \href {http://dx.doi.org/#2} {#1}\fi
  \endgroup}
\def\mn@eprint#1#2{\mn@eprint@#1:#2::\@nil}
\def\mn@eprint@arXiv#1{\href {http://arxiv.org/abs/#1} {{\tt arXiv:#1}}}
\def\mn@eprint@dblp#1{\href {http://dblp.uni-trier.de/rec/bibtex/#1.xml}
  {dblp:#1}}
\def\mn@eprint@#1:#2:#3:#4\@nil{\def\@tempa {#1}\def\@tempb {#2}\def\@tempc
  {#3}\ifx \@tempc \@empty \let \@tempc \@tempb \let \@tempb \@tempa \fi \ifx
  \@tempb \@empty \def\@tempb {arXiv}\fi \@ifundefined
  {mn@eprint@\@tempb}{\@tempb:\@tempc}{\expandafter \expandafter \csname
  mn@eprint@\@tempb\endcsname \expandafter{\@tempc}}}

\bibitem[\protect\citeauthoryear{{Arentsen} et~al.,}{{Arentsen}
  et~al.}{2020}]{arentsen2020}
{Arentsen} A.,  et~al., 2020, \mn@doi [\mnras] {10.1093/mnras/staa1661}, \href
  {https://ui.adsabs.harvard.edu/abs/2020MNRAS.496.4964A} {496, 4964}

\bibitem[\protect\citeauthoryear{{Badenes} et~al.,}{{Badenes}
  et~al.}{2018}]{badenes18}
{Badenes} C.,  et~al., 2018, \mn@doi [\apj] {10.3847/1538-4357/aaa765}, \href
  {https://ui.adsabs.harvard.edu/abs/2018ApJ...854..147B} {854, 147}

\bibitem[\protect\citeauthoryear{{Battaglia}, {Taibi}, {Thomas}  \&
  {Fritz}}{{Battaglia} et~al.}{2022}]{battaglia2022}
{Battaglia} G.,  {Taibi} S.,  {Thomas} G.~F.,   {Fritz} T.~K.,  2022, \mn@doi
  [\aap] {10.1051/0004-6361/202141528}, \href
  {https://ui.adsabs.harvard.edu/abs/2022A&A...657A..54B} {657, A54}

\bibitem[\protect\citeauthoryear{{Bullock} \& {Boylan-Kolchin}}{{Bullock} \&
  {Boylan-Kolchin}}{2017}]{bullock2017}
{Bullock} J.~S.,  {Boylan-Kolchin} M.,  2017, \mn@doi [\araa]
  {10.1146/annurev-astro-091916-055313}, \href
  {https://ui.adsabs.harvard.edu/abs/2017ARA&A..55..343B} {55, 343}

\bibitem[\protect\citeauthoryear{{Carlin} et~al.,}{{Carlin}
  et~al.}{2017}]{carlin2017}
{Carlin} J.~L.,  et~al., 2017, \mn@doi [\aj] {10.3847/1538-3881/aa94d0}, \href
  {https://ui.adsabs.harvard.edu/abs/2017AJ....154..267C} {154, 267}

\bibitem[\protect\citeauthoryear{{Choi}, {Dotter}, {Conroy}, {Cantiello},
  {Paxton}  \& {Johnson}}{{Choi} et~al.}{2016}]{MISTisochrone1}
{Choi} J.,  {Dotter} A.,  {Conroy} C.,  {Cantiello} M.,  {Paxton} B.,
  {Johnson} B.~D.,  2016, \mn@doi [\apj] {10.3847/0004-637X/823/2/102}, \href
  {https://ui.adsabs.harvard.edu/abs/2016ApJ...823..102C} {823, 102}

\bibitem[\protect\citeauthoryear{{Cooper}, {Newman}, {Davis}, {Finkbeiner}  \&
  {Gerke}}{{Cooper} et~al.}{2012}]{cooper2012}
{Cooper} M.~C.,  {Newman} J.~A.,  {Davis} M.,  {Finkbeiner} D.~P.,   {Gerke}
  B.~F.,  2012, {spec2d: DEEP2 DEIMOS Spectral Pipeline} (\mn@eprint {ascl}
  {1203.003})

\bibitem[\protect\citeauthoryear{{Dotter}}{{Dotter}}{2016}]{MISTisochrone0}
{Dotter} A.,  2016, \mn@doi [\apjs] {10.3847/0067-0049/222/1/8}, \href
  {https://ui.adsabs.harvard.edu/abs/2016ApJS..222....8D} {222, 8}

\bibitem[\protect\citeauthoryear{{Errani}, {Pe{\~n}arrubia}  \&
  {Walker}}{{Errani} et~al.}{2018}]{errani2018}
{Errani} R.,  {Pe{\~n}arrubia} J.,   {Walker} M.~G.,  2018, \mn@doi [\mnras]
  {10.1093/mnras/sty2505}, \href
  {https://ui.adsabs.harvard.edu/abs/2018MNRAS.481.5073E} {481, 5073}

\bibitem[\protect\citeauthoryear{{Faber} et~al.,}{{Faber}
  et~al.}{2003}]{faber2003}
{Faber} S.~M.,  et~al., 2003, in {Iye} M.,  {Moorwood} A. F.~M.,  eds,  Society
  of Photo-Optical Instrumentation Engineers (SPIE) Conference Series Vol.
  4841, Instrument Design and Performance for Optical/Infrared Ground-based
  Telescopes. pp 1657--1669, \mn@doi{10.1117/12.460346}

\bibitem[\protect\citeauthoryear{{Flewelling} et~al.,}{{Flewelling}
  et~al.}{2020}]{pan-starrs1_cite}
{Flewelling} H.~A.,  et~al., 2020, \mn@doi [\apjs] {10.3847/1538-4365/abb82d},
  \href {https://ui.adsabs.harvard.edu/abs/2020ApJS..251....7F} {251, 7}

\bibitem[\protect\citeauthoryear{{Foreman-Mackey}, {Hogg}, {Lang}  \&
  {Goodman}}{{Foreman-Mackey} et~al.}{2013}]{foreman-mackey2013}
{Foreman-Mackey} D.,  {Hogg} D.~W.,  {Lang} D.,   {Goodman} J.,  2013, \mn@doi
  [\pasp] {10.1086/670067}, \href
  {https://ui.adsabs.harvard.edu/abs/2013PASP..125..306F} {125, 306}

\bibitem[\protect\citeauthoryear{{Gaia Collaboration} et~al.,}{{Gaia
  Collaboration} et~al.}{2021}]{gaiaedr3}
{Gaia Collaboration} et~al., 2021, \mn@doi [\aap]
  {10.1051/0004-6361/202039657}, \href
  {https://ui.adsabs.harvard.edu/abs/2021A&A...649A...1G} {649, A1}

\bibitem[\protect\citeauthoryear{{Hekker}, {Snellen}, {Aerts}, {Quirrenbach},
  {Reffert}  \& {Mitchell}}{{Hekker} et~al.}{2008}]{hekker2008}
{Hekker} S.,  {Snellen} I.~A.~G.,  {Aerts} C.,  {Quirrenbach} A.,  {Reffert}
  S.,   {Mitchell} D.~S.,  2008, in Journal of Physics Conference Series. p.
  012058 (\mn@eprint {arXiv} {0710.4134}),
  \mn@doi{10.1088/1742-6596/118/1/012058}

\bibitem[\protect\citeauthoryear{{Ibata}, {Sollima}, {Nipoti}, {Bellazzini},
  {Chapman}  \& {Dalessandro}}{{Ibata} et~al.}{2011}]{ibata2011}
{Ibata} R.,  {Sollima} A.,  {Nipoti} C.,  {Bellazzini} M.,  {Chapman} S.~C.,
  {Dalessandro} E.,  2011, \mn@doi [\apj] {10.1088/0004-637X/738/2/186}, \href
  {https://ui.adsabs.harvard.edu/abs/2011ApJ...738..186I} {738, 186}

\bibitem[\protect\citeauthoryear{{Illingworth}}{{Illingworth}}{1976}]{illingworth76}
{Illingworth} G.,  1976, \mn@doi [\apj] {10.1086/154152}, \href
  {https://ui.adsabs.harvard.edu/abs/1976ApJ...204...73I} {204, 73}

\bibitem[\protect\citeauthoryear{{Irwin} \& {Lewis}}{{Irwin} \&
  {Lewis}}{2001}]{irwin2001}
{Irwin} M.,  {Lewis} J.,  2001, \mn@doi [\nar] {10.1016/S1387-6473(00)00138-X},
  \href {https://ui.adsabs.harvard.edu/abs/2001NewAR..45..105I} {45, 105}

\bibitem[\protect\citeauthoryear{{Ji}, {Simon}, {Frebel}, {Venn}  \&
  {Hansen}}{{Ji} et~al.}{2020}]{ji2020}
{Ji} A.~P.,  {Simon} J.~D.,  {Frebel} A.,  {Venn} K.~A.,   {Hansen} T.~T.,
  2020, VizieR Online Data Catalog, \href
  {https://ui.adsabs.harvard.edu/abs/2020yCat..18700083J} {p. J/ApJ/870/83}

\bibitem[\protect\citeauthoryear{{Kirby}, {Cohen}, {Simon}  \&
  {Guhathakurta}}{{Kirby} et~al.}{2015}]{kirby2015}
{Kirby} E.~N.,  {Cohen} J.~G.,  {Simon} J.~D.,   {Guhathakurta} P.,  2015,
  \mn@doi [\apjl] {10.1088/2041-8205/814/1/L7}, \href
  {https://ui.adsabs.harvard.edu/abs/2015ApJ...814L...7K} {814, L7}

\bibitem[\protect\citeauthoryear{{Kirby}, {Cohen}, {Simon}, {Guhathakurta},
  {Thygesen}  \& {Duggan}}{{Kirby} et~al.}{2017}]{kirby2017}
{Kirby} E.~N.,  {Cohen} J.~G.,  {Simon} J.~D.,  {Guhathakurta} P.,  {Thygesen}
  A.~O.,   {Duggan} G.~E.,  2017, \mn@doi [\apj] {10.3847/1538-4357/aa6570},
  \href {https://ui.adsabs.harvard.edu/abs/2017ApJ...838...83K} {838, 83}

\bibitem[\protect\citeauthoryear{{Koch}, {Hansen}, {Feltzing}  \&
  {Wilkinson}}{{Koch} et~al.}{2014}]{koch2014}
{Koch} A.,  {Hansen} T.,  {Feltzing} S.,   {Wilkinson} M.~I.,  2014, \mn@doi
  [\apj] {10.1088/0004-637X/780/1/91}, \href
  {https://ui.adsabs.harvard.edu/abs/2014ApJ...780...91K} {780, 91}

\bibitem[\protect\citeauthoryear{{Koposov} et~al.,}{{Koposov}
  et~al.}{2011}]{koposov11}
{Koposov} S.~E.,  et~al., 2011, \mn@doi [\apj] {10.1088/0004-637X/736/2/146},
  \href {https://ui.adsabs.harvard.edu/abs/2011ApJ...736..146K} {736, 146}

\bibitem[\protect\citeauthoryear{Laevens et~al.,}{Laevens
  et~al.}{2015}]{Laevens2015}
Laevens B. P.~M.,  et~al., 2015, \mn@doi [The Astrophysical Journal]
  {10.1088/2041-8205/802/2/l18}, 802, L18

\bibitem[\protect\citeauthoryear{Li et~al.,}{Li et~al.}{2019}]{li2019}
Li T.~S.,  et~al., 2019, \mn@doi [Monthly Notices of the Royal Astronomical
  Society] {10.1093/mnras/stz2731}, 490, 3508

\bibitem[\protect\citeauthoryear{{Magnier} \& {Cuillandre}}{{Magnier} \&
  {Cuillandre}}{2004}]{Magnier2004}
{Magnier} E.~A.,  {Cuillandre} J.~C.,  2004, \mn@doi [\pasp] {10.1086/420756},
  \href {https://ui.adsabs.harvard.edu/abs/2004PASP..116..449M} {116, 449}

\bibitem[\protect\citeauthoryear{Martin et~al.,}{Martin
  et~al.}{2016}]{martin2016}
Martin N.~F.,  et~al., 2016, \mn@doi [The Astrophysical Journal]
  {10.3847/0004-637x/818/1/40}, 818, 40

\bibitem[\protect\citeauthoryear{{Martinez}, {Minor}, {Bullock}, {Kaplinghat},
  {Simon}  \& {Geha}}{{Martinez} et~al.}{2011}]{martinez2011}
{Martinez} G.~D.,  {Minor} Q.~E.,  {Bullock} J.,  {Kaplinghat} M.,  {Simon}
  J.~D.,   {Geha} M.,  2011, \mn@doi [\apj] {10.1088/0004-637X/738/1/55}, \href
  {https://ui.adsabs.harvard.edu/abs/2011ApJ...738...55M} {738, 55}

\bibitem[\protect\citeauthoryear{{Mateo}}{{Mateo}}{1998}]{mateo98}
{Mateo} M.~L.,  1998, \mn@doi [\araa] {10.1146/annurev.astro.36.1.435}, \href
  {https://ui.adsabs.harvard.edu/abs/1998ARA&A..36..435M} {36, 435}

\bibitem[\protect\citeauthoryear{{Mazzola} et~al.,}{{Mazzola}
  et~al.}{2020}]{mazzola2020}
{Mazzola} C.~N.,  et~al., 2020, \mn@doi [\mnras] {10.1093/mnras/staa2859},
  \href {https://ui.adsabs.harvard.edu/abs/2020MNRAS.499.1607M} {499, 1607}

\bibitem[\protect\citeauthoryear{{McConnachie}}{{McConnachie}}{2012}]{mcconnachie12}
{McConnachie} A.~W.,  2012, \mn@doi [\aj] {10.1088/0004-6256/144/1/4}, \href
  {https://ui.adsabs.harvard.edu/abs/2012AJ....144....4M} {144, 4}

\bibitem[\protect\citeauthoryear{{McConnachie} \& {C{\^o}t{\'e}}}{{McConnachie}
  \& {C{\^o}t{\'e}}}{2010}]{mcconnachie2010}
{McConnachie} A.~W.,  {C{\^o}t{\'e}} P.,  2010, \mn@doi [\apjl]
  {10.1088/2041-8205/722/2/L209}, \href
  {https://ui.adsabs.harvard.edu/abs/2010ApJ...722L.209M} {722, L209}

\bibitem[\protect\citeauthoryear{McConnachie \& Venn}{McConnachie \&
  Venn}{2020}]{mcconnachie2020updated}
McConnachie A.~W.,  Venn K.~A.,  2020, Updated proper motions for Local Group
  dwarf galaxies using Gaia Early Data Release 3 (\mn@eprint {arXiv}
  {2012.03904})

\bibitem[\protect\citeauthoryear{{Minor}}{{Minor}}{2013}]{minor2013}
{Minor} Q.~E.,  2013, \mn@doi [\apj] {10.1088/0004-637X/779/2/116}, \href
  {https://ui.adsabs.harvard.edu/abs/2013ApJ...779..116M} {779, 116}

\bibitem[\protect\citeauthoryear{{Minor}, {Martinez}, {Bullock}, {Kaplinghat}
  \& {Trainor}}{{Minor} et~al.}{2010}]{Minor2010ApJ...721.1142M}
{Minor} Q.~E.,  {Martinez} G.,  {Bullock} J.,  {Kaplinghat} M.,   {Trainor} R.,
   2010, \mn@doi [\apj] {10.1088/0004-637X/721/2/1142}, \href
  {http://adsabs.harvard.edu/abs/2010ApJ...721.1142M} {721, 1142}

\bibitem[\protect\citeauthoryear{{Minor}, {Pace}, {Marshall}  \&
  {Strigari}}{{Minor} et~al.}{2019}]{Minor2019MNRAS.487.2961M}
{Minor} Q.~E.,  {Pace} A.~B.,  {Marshall} J.~L.,   {Strigari} L.~E.,  2019,
  \mn@doi [\mnras] {10.1093/mnras/stz1468}, \href
  {https://ui.adsabs.harvard.edu/abs/2019MNRAS.487.2961M} {487, 2961}

\bibitem[\protect\citeauthoryear{{Miyazaki} et~al.,}{{Miyazaki}
  et~al.}{2018}]{HSC_cite}
{Miyazaki} S.,  et~al., 2018, \mn@doi [\pasj] {10.1093/pasj/psx063}, \href
  {https://ui.adsabs.harvard.edu/abs/2018PASJ...70S...1M} {70, S1}

\bibitem[\protect\citeauthoryear{{Moe}, {Kratter}  \& {Badenes}}{{Moe}
  et~al.}{2019}]{moe2019}
{Moe} M.,  {Kratter} K.~M.,   {Badenes} C.,  2019, \mn@doi [\apj]
  {10.3847/1538-4357/ab0d88}, \href
  {https://ui.adsabs.harvard.edu/abs/2019ApJ...875...61M} {875, 61}

\bibitem[\protect\citeauthoryear{{Morton}}{{Morton}}{2015}]{morton2015}
{Morton} T.~D.,  2015, {isochrones: Stellar model grid package} (\mn@eprint
  {ascl} {1503.010})

\bibitem[\protect\citeauthoryear{{Murray} \& {Correia}}{{Murray} \&
  {Correia}}{2010}]{murrayandcorreia2010}
{Murray} C.~D.,  {Correia} A.~C.~M.,  2010, {Keplerian Orbits and Dynamics of
  Exoplanets}.
pp 15--23

\bibitem[\protect\citeauthoryear{{Oke} et~al.,}{{Oke} et~al.}{1995}]{oke1995}
{Oke} J.~B.,  et~al., 1995, \mn@doi [\pasp] {10.1086/133562}, \href
  {https://ui.adsabs.harvard.edu/abs/1995PASP..107..375O} {107, 375}

\bibitem[\protect\citeauthoryear{{Olszewski}, {Pryor}  \&
  {Armandroff}}{{Olszewski} et~al.}{1996}]{Olszewski1996}
{Olszewski} E.~W.,  {Pryor} C.,   {Armandroff} T.~E.,  1996, \mn@doi [\aj]
  {10.1086/117821}, \href
  {https://ui.adsabs.harvard.edu/abs/1996AJ....111..750O} {111, 750}

\bibitem[\protect\citeauthoryear{{Pace} et~al.,}{{Pace}
  et~al.}{2020}]{pace2020}
{Pace} A.~B.,  et~al., 2020, \mn@doi [\mnras] {10.1093/mnras/staa1419}, \href
  {https://ui.adsabs.harvard.edu/abs/2020MNRAS.495.3022P} {495, 3022}

\bibitem[\protect\citeauthoryear{{Pace}, {Walker}, {Koposov}, {Caldwell},
  {Mateo}, {Olszewski}, {Bailey}  \& {Wang}}{{Pace} et~al.}{2021}]{pace2021}
{Pace} A.~B.,  {Walker} M.~G.,  {Koposov} S.~E.,  {Caldwell} N.,  {Mateo} M.,
  {Olszewski} E.~W.,  {Bailey} John~I. I.,   {Wang} M.-Y.,  2021, arXiv
  e-prints, \href {https://ui.adsabs.harvard.edu/abs/2021arXiv210500064P} {p.
  arXiv:2105.00064}

\bibitem[\protect\citeauthoryear{{Pace}, {Erkal}  \& {Li}}{{Pace}
  et~al.}{2022}]{Pace2022arXiv220505699P}
{Pace} A.~B.,  {Erkal} D.,   {Li} T.~S.,  2022, arXiv e-prints, \href
  {https://ui.adsabs.harvard.edu/abs/2022arXiv220505699P} {p. arXiv:2205.05699}

\bibitem[\protect\citeauthoryear{Pedregosa et~al.,}{Pedregosa
  et~al.}{2011}]{scikit-learn}
Pedregosa F.,  et~al., 2011, Journal of Machine Learning Research, 12, 2825

\bibitem[\protect\citeauthoryear{{Price-Whelan}, {Hogg}, {Foreman-Mackey}  \&
  {Rix}}{{Price-Whelan} et~al.}{2017}]{price-whelan2017}
{Price-Whelan} A.~M.,  {Hogg} D.~W.,  {Foreman-Mackey} D.,   {Rix} H.-W.,
  2017, \mn@doi [\apj] {10.3847/1538-4357/aa5e50}, \href
  {https://ui.adsabs.harvard.edu/abs/2017ApJ...837...20P} {837, 20}

\bibitem[\protect\citeauthoryear{{Raghavan} et~al.,}{{Raghavan}
  et~al.}{2010}]{raghavan2010}
{Raghavan} D.,  et~al., 2010, \mn@doi [\apjs] {10.1088/0067-0049/190/1/1},
  \href {https://ui.adsabs.harvard.edu/abs/2010ApJS..190....1R} {190, 1}

\bibitem[\protect\citeauthoryear{{Simon}}{{Simon}}{2019}]{simon2019}
{Simon} J.~D.,  2019, \mn@doi [\araa] {10.1146/annurev-astro-091918-104453},
  \href {https://ui.adsabs.harvard.edu/abs/2019ARA&A..57..375S} {57, 375}

\bibitem[\protect\citeauthoryear{{Sohn} et~al.,}{{Sohn}
  et~al.}{2007}]{sohn2007}
{Sohn} S.~T.,  et~al., 2007, \mn@doi [\apj] {10.1086/518302}, \href
  {http://adsabs.harvard.edu/abs/2007ApJ...663..960S} {663, 960}

\bibitem[\protect\citeauthoryear{{Spencer}, {Mateo}, {Walker}, {Olszewski},
  {McConnachie}, {Kirby}  \& {Koch}}{{Spencer} et~al.}{2017}]{spencer2017}
{Spencer} M.~E.,  {Mateo} M.,  {Walker} M.~G.,  {Olszewski} E.~W.,
  {McConnachie} A.~W.,  {Kirby} E.~N.,   {Koch} A.,  2017, \mn@doi [\aj]
  {10.3847/1538-3881/aa6d51}, \href
  {https://ui.adsabs.harvard.edu/abs/2017AJ....153..254S} {153, 254}

\bibitem[\protect\citeauthoryear{{Spencer}, {Mateo}, {Olszewski}, {Walker},
  {McConnachie}  \& {Kirby}}{{Spencer} et~al.}{2018}]{spencer2018}
{Spencer} M.~E.,  {Mateo} M.,  {Olszewski} E.~W.,  {Walker} M.~G.,
  {McConnachie} A.~W.,   {Kirby} E.~N.,  2018, \mn@doi [\aj]
  {10.3847/1538-3881/aae3e4}, \href
  {https://ui.adsabs.harvard.edu/abs/2018AJ....156..257S} {156, 257}

\bibitem[\protect\citeauthoryear{{Starkenburg} et~al.,}{{Starkenburg}
  et~al.}{2017}]{starkenburg2017}
{Starkenburg} E.,  et~al., 2017, \mn@doi [\mnras] {10.1093/mnras/stx1068},
  \href {https://ui.adsabs.harvard.edu/abs/2017MNRAS.471.2587S} {471, 2587}

\bibitem[\protect\citeauthoryear{{Szentgyorgyi} et~al.,}{{Szentgyorgyi}
  et~al.}{2011}]{szentgyorgyi2011}
{Szentgyorgyi} A.,  et~al., 2011, \mn@doi [\pasp] {10.1086/662209}, \href
  {https://ui.adsabs.harvard.edu/abs/2011PASP..123.1188S} {123, 1188}

\bibitem[\protect\citeauthoryear{{Tolstoy}, {Hill}  \& {Tosi}}{{Tolstoy}
  et~al.}{2009}]{tolstoy09}
{Tolstoy} E.,  {Hill} V.,   {Tosi} M.,  2009, \mn@doi [\araa]
  {10.1146/annurev-astro-082708-101650}, \href
  {https://ui.adsabs.harvard.edu/abs/2009ARA&A..47..371T} {47, 371}

\bibitem[\protect\citeauthoryear{{Venn}, {Starkenburg}, {Malo}, {Martin}  \&
  {Laevens}}{{Venn} et~al.}{2017}]{venn2017}
{Venn} K.~A.,  {Starkenburg} E.,  {Malo} L.,  {Martin} N.,   {Laevens}
  B.~P.~M.,  2017, \mn@doi [\mnras] {10.1093/mnras/stw3198}, \href
  {https://ui.adsabs.harvard.edu/abs/2017MNRAS.466.3741V} {466, 3741}

\bibitem[\protect\citeauthoryear{{Walker}, {Mateo}, {Olszewski},
  {Pe{\~n}arrubia}, {Evans}  \& {Gilmore}}{{Walker} et~al.}{2009}]{walker09}
{Walker} M.~G.,  {Mateo} M.,  {Olszewski} E.~W.,  {Pe{\~n}arrubia} J.,  {Evans}
  N.~W.,   {Gilmore} G.,  2009, \mn@doi [\apj] {10.1088/0004-637X/704/2/1274},
  \href {https://ui.adsabs.harvard.edu/abs/2009ApJ...704.1274W} {704, 1274}

\bibitem[\protect\citeauthoryear{{Walker}, {Olszewski}  \& {Mateo}}{{Walker}
  et~al.}{2015}]{walker15draco}
{Walker} M.~G.,  {Olszewski} E.~W.,   {Mateo} M.,  2015, \mn@doi [\mnras]
  {10.1093/mnras/stv099}, \href
  {https://ui.adsabs.harvard.edu/abs/2015MNRAS.448.2717W} {448, 2717}

\bibitem[\protect\citeauthoryear{{Willman} \& {Strader}}{{Willman} \&
  {Strader}}{2012}]{willman_strader2012}
{Willman} B.,  {Strader} J.,  2012, \mn@doi [\aj] {10.1088/0004-6256/144/3/76},
  \href {https://ui.adsabs.harvard.edu/abs/2012AJ....144...76W} {144, 76}

\bibitem[\protect\citeauthoryear{{Wolf}, {Martinez}, {Bullock}, {Kaplinghat},
  {Geha}, {Mu{\~n}oz}, {Simon}  \& {Avedo}}{{Wolf} et~al.}{2010}]{wolf10}
{Wolf} J.,  {Martinez} G.~D.,  {Bullock} J.~S.,  {Kaplinghat} M.,  {Geha} M.,
  {Mu{\~n}oz} R.~R.,  {Simon} J.~D.,   {Avedo} F.~F.,  2010, \mn@doi [\mnras]
  {10.1111/j.1365-2966.2010.16753.x}, \href
  {https://ui.adsabs.harvard.edu/abs/2010MNRAS.406.1220W} {406, 1220}

\makeatother
\end{thebibliography}




\appendix

\section{Lower limit on $\log_{10}\sigma_v$ prior}\label{appendix: sigv_prior}
From the virial theorem, we know that the mass contained within the half-light radius  of a dwarf galaxy in virial equilibrium can be described by...
\begin{equation}
    M (< R_{1/2}) = C \frac{R_{1/2} \sigma_v^2}{G}
\end{equation}

where $\sigma_v$ is the galaxy's velocity dispersion and $G$ is the gravitational constant. $C$ is a proportionality constant. Assuming that all of the mass is accounted for by the total stellar mass, the following relation becomes true.
\begin{equation}
    M (< R_{1/2}) = \frac{ \Upsilon L }{2}
\end{equation}
where $\Upsilon$ is the stellar mass-to-light ratio and L is the total luminosity. Combining these, we can solve for $\sigma_v$.
\begin{equation}
    \sigma_v = \sqrt{\frac{G}{2C} \frac{\Upsilon L}{R_{1/2}}}
\end{equation}

It is estimated that Tri~II has a total luminosity $L \sim 400 \rm~L_\odot$ and a half-light radius $R_{1/2} \sim 34 \rm~pc$ \citep{Laevens2015}. Stellar mass to light ratios usually exist in the rage $\Upsilon \sim 0.5$ to $3.0 \rm~M_\odot~L_\odot^{-1}$ and the proportionality constant  is in the range $C \sim 2$ to $4$. Taking the lower limit for $\Upsilon$ and the upper limit for $C$, we find that $\sigma_v$ can go as small as $0.05 \rm~km~s^{-1}$. This value will serve as the lower limit of our velocity dispersion prior and corresponds to a system where the dynamics are fully determined by visible matter.

\section{GRACES}\label{appendix: GRACES}

\begin{figure}
    \centering
    \includegraphics[width=\columnwidth]{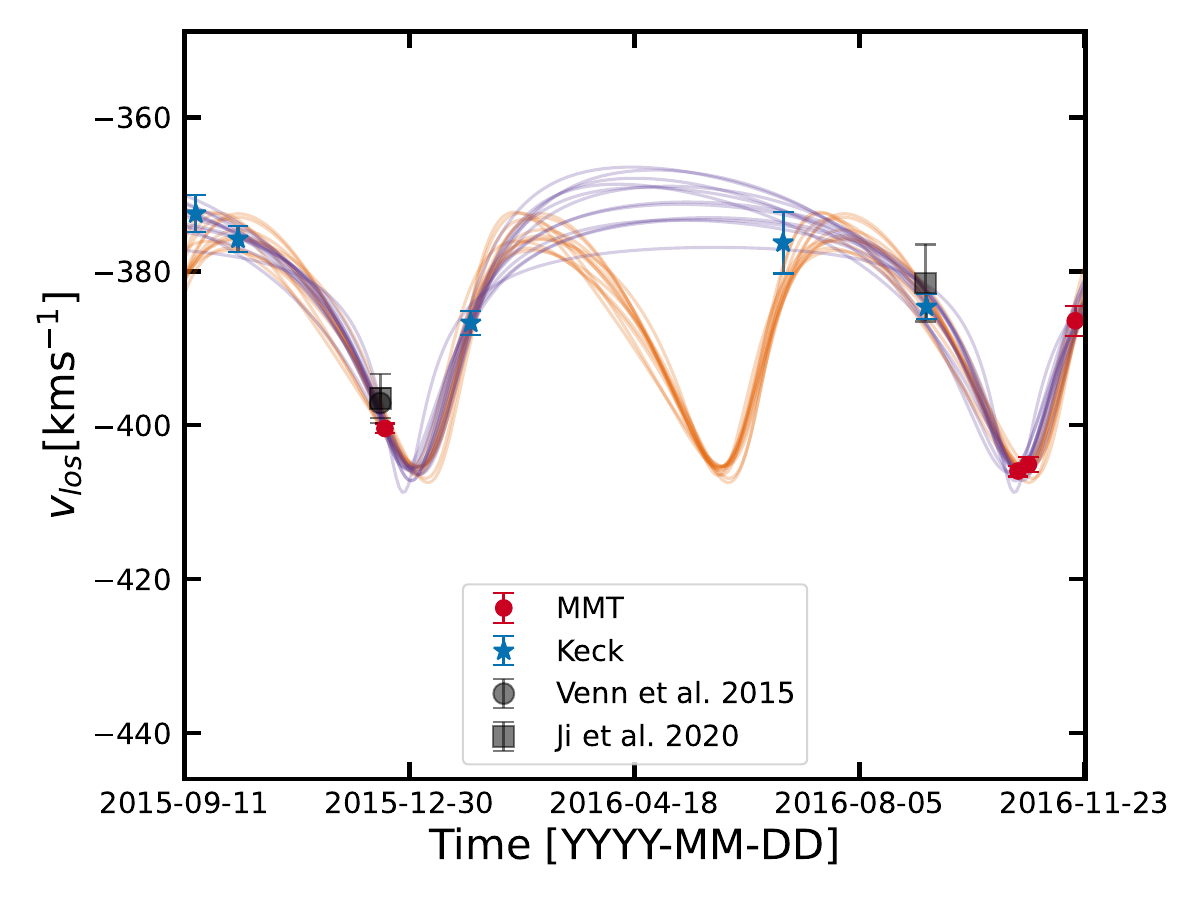}
    \caption{Radial velocity curve of {\it Star46} including GRACES epochs. The \citet{venn2017} epoch and the first \citet{ji2020} epoch overlap. The orbital solutions plotted are the results of finding orbital solutions  to the inclusion of the two \citet{ji2020} observations. The orange and purple lines correspond to solutions with orbital periods of approximately 0.4 years and 0.8 years respectively.} 
    \label{fig:binary_rv_graces}
\end{figure}

We briefly explored how the inclusion of the {\it Star46} measurements in \cite{venn2017} and \cite{ji2020} would impact the binary orbital parameter posterior. For this, we do not apply an offset to the velocities measured in GRACES because of the lack of overlap in observation fields between GRACES and either MMT or Keck. Figure \ref{fig:binary_rv_graces} shows the new epochs relative to the radial velocity curve presented in the main text (see Table \ref{tab:GRACES} for the individual epoch measurements).

The spectra taken in December 2015 are reduced using different pipelines in the two papers, resulting in slightly different but consistent measured velocities for the same epoch (See Section 2 of \citealt[][]{venn2017} and Section 2 of \citealt[][]{ji2020}).

\begin{table}
    \centering
    \begin{tabular}{c|c|c}
         {\bf Source } & {\bf HJD} &  {\bf$v_{\rm los}$ [$\rm km~s^{-1}$]} \\
         \hline \hline
         Venn et al. 2017 & 2457372.5 & $-397.1 \pm 2.0$\\
         \hline
         Ji et al. 2020 & 2457372.5 & $-396.5 \pm 3.2$ \\
         Ji et al. 2020 & 2457638.5 & $-381.5 \pm 5.0$\\
    \end{tabular}
    \caption{Additional GRACES observations of the Tri~II binary, {\it Star46}.}
    \label{tab:GRACES}
\end{table}

\begin{figure}
    \centering
    \includegraphics[width=\columnwidth]{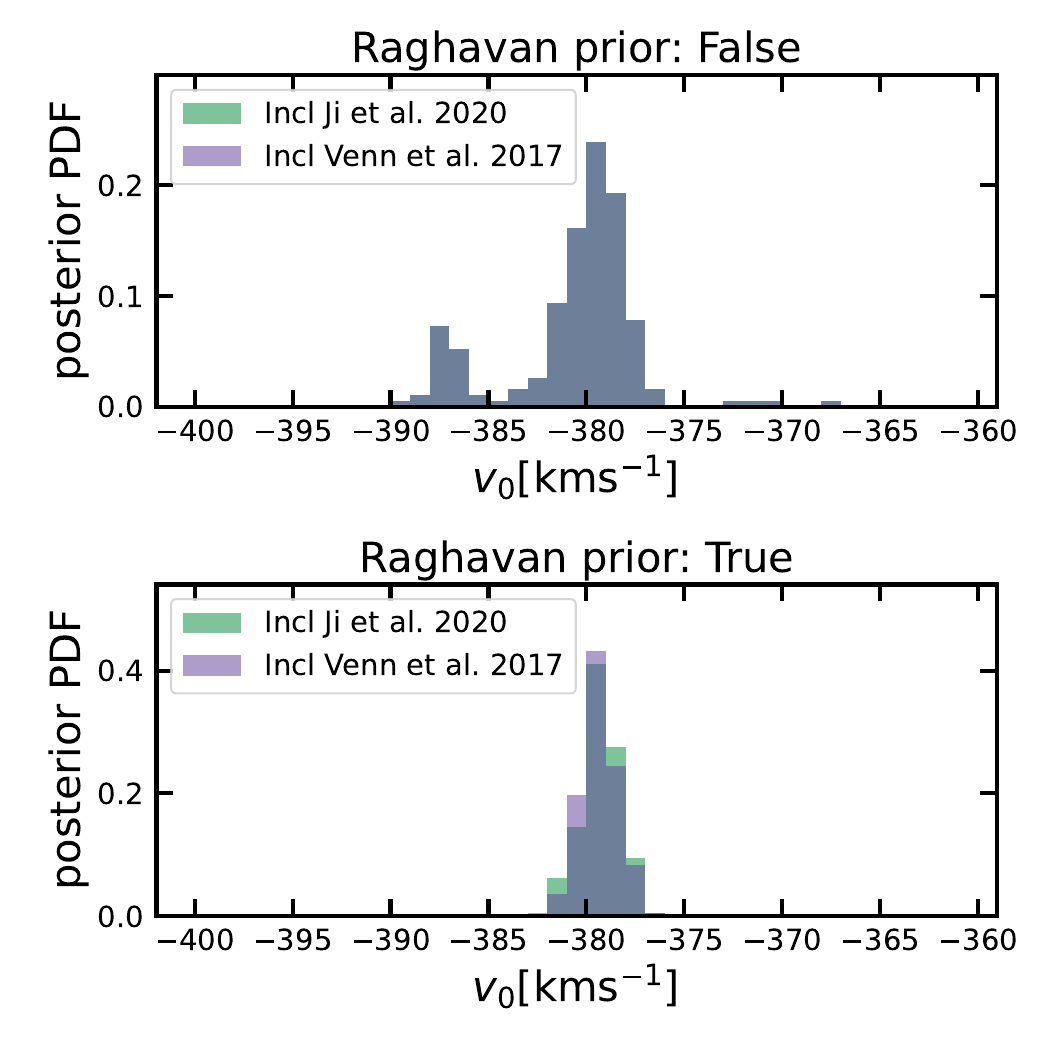}
    \caption{Resulting systemic velocity posteriors for {\it Star46} when including GRACES epochs. The posteriors from including  \citet{venn2017}and \citet{ji2020} are over plotted in different colors. The overlap in the histograms is colored grey.}
    \label{fig:v0_graces}
\end{figure}

Figure \ref{fig:v0_graces} shows the inferred systemic velocity posterior when including GRACES observations. Under the default period prior, the posterior takes on the same bimodal shape as seen in Section \ref{sec:orbital_params} regardless of which set of GRACES observations is used. Compared to the posteriors of the systemic velocity found in the main text, with modes at $-387.6_{-1.2}^{+0.8}$ $\rm~km~s^{-1}$ and $-380.0_{-1.7}^{+1.8}$ $\rm~km~s^{-1}$, there is a slight shift in the long-period systemic velocity. The new corresponding modes are at $-387.18\pm_{0.69}^{0.81}$ $\rm~km~s^{-1}$ and $-379.51\pm_{1.51}^{1.32}$ $\rm~km~s^{-1}$ when including either set of GRACES observations. Applying the period distribution found in \citet{raghavan2010}, there is again re-weighting toward the long-period solutions. There is a very small disparity between the resulting unimodal posteriors with the systemic velocity becoming $-379.48\pm_{0.87}^{1.14}$ when using \cite{ji2020}'s observations and $-379.38\pm_{1.22}^{0.98}$ when using\cite{venn2017}'s observation. However, the differences of these posteriors from those in the main text are well within error.

While these new values have slightly smaller errors, this improvement may likely vanish after properly accounting for any zero-point offsets between GRACES and MMT or Keck. These is also not a clear conclusion of which of the two GRACES sets is preferable for this analysis. Rather than exploring this avenue that would likely yield only a marginal improvement, we opt to not include the GRACES observations of this star in this paper.

\section{Metallicity Dispersion}\label{appendix: feh_disp}
\begin{figure}
    \centering
    \includegraphics[width=\columnwidth]{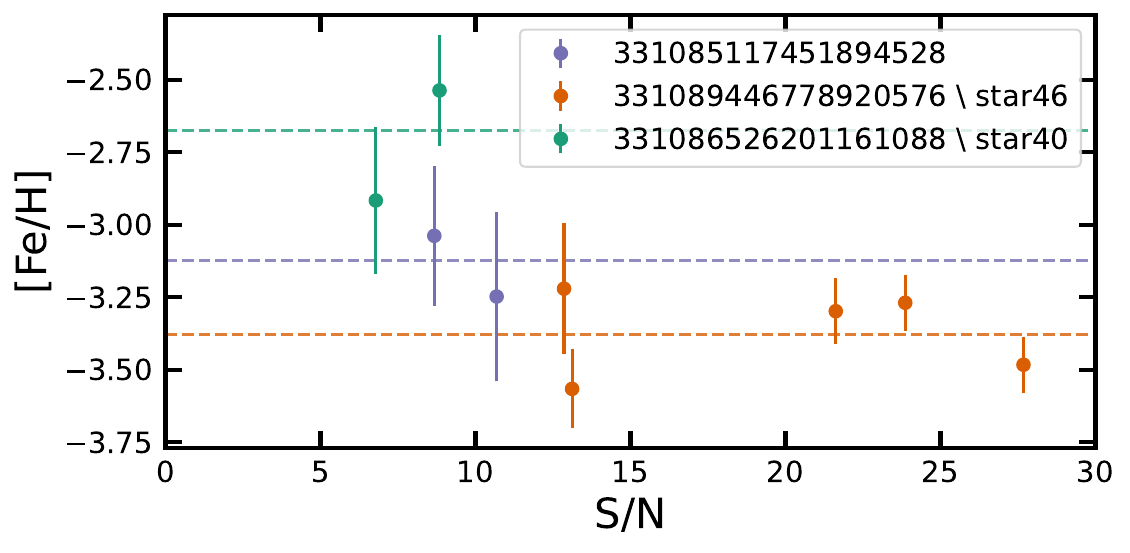}
    \caption{Fe/H measurements v. signal-to-noise of the Tri~II members in our MMT catalog after a quality cut, colored by object. The dotted lines mark the weighted mean Fe/H values for each star.}
    \label{fig:MMT_FeH_SN}
\end{figure}

In this section, we briefly explore how the addition of the new MMT data affects the metalicity disperison of Tri~II. From repeat measurements in the MMT catalog, we only consider measurements with an Fe/H error < 0.4 dex as being good-quality observations. At lower signal-to-noise, we see a bias for higher metallicities in dwarf galaxy members and applying this selection removes those bad measurements. We follow the procedure performed in  \cite[Section 4 of ][]{kirby2017} to calculate the metallicity dispersion, performing maximum likelihood estimation to fit a Gaussian distribution to the Fe/H measurements of Tri~II members. We assume the existence of a Fe/H zero-point offset between the instruments which was determined using the non-outlier offset model (setting outlier fraction to zero in equation \ref{eq:outlier_like_model}) with the two stars that are in the overlap of the MMT and Keck catalogues, {\it Star46} and {\it Star40}. We find an offset of $-0.63 \pm 0.20 \rm~dex$ to be added to the \cite{kirby2017} metallicities. We note that a weighted mean is not the preferred way of combining metallicity measurements and improvements to signal-to-noise would be achieved by co-adding the relevant spectra. Figure \ref{fig:MMT_FeH_SN} shows Fe/H measurements for the stars observed in MMT.

\begin{figure}
    \centering
    \includegraphics[width=\columnwidth]{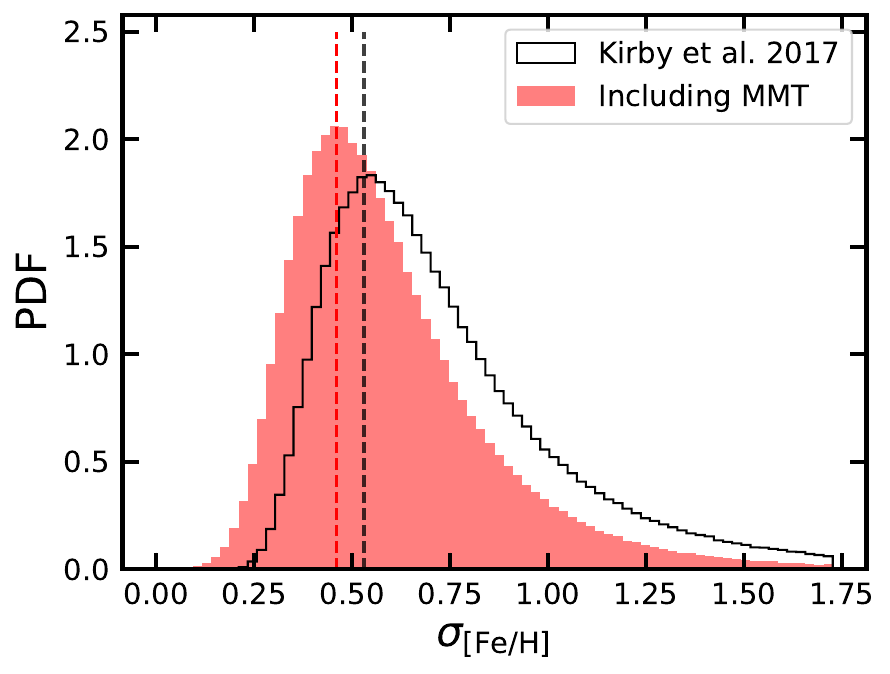}
    \caption{Metallicity dispersion likelihood distributions calculated when using only the Keck data \citep{kirby2017} for Tri~II (black) and when also including our Tri~II MMT measurements (red). The dotted lines mark the maximum likelihood values.}
    \label{fig:sig_FeH}
\end{figure}

\cite{kirby2017} found $\sigma([\rm Fe/H]) = 0.53^{+0.38}_{-0.12} \rm~dex$ when including all potential member stars in their catalog \citep[Figure 6b of ][]{kirby2017}. The effect of including the MMT observations can be seen in Figure \ref{fig:sig_FeH}, slighlty shifting the maximum likelihood value to $\sigma([\rm Fe/H]) = 0.46_{-0.09}^{+0.37}$.

\section{Data Tables}\label{appendix: data_tables}

\clearpage
\begin{sidewaystable}
\centering

\begin{tabular}{cccccccccccc}
Gaia ID & KCS2015 & MIC2016 & Member & RA (ICRS)  & Dec (ICRS) & HJD (days) & S/N & $v_{\rm los} (\rm km~s^{-1})$ & [Fe/H]  & $\rm T_{eff}$ & $\log g$\\ \hline     \hline
 331085117451894528 &  &  & True & 33.4272917 & 36.0950281 & 2457683.72 & 8.67 & -386.9$\pm$0.7 & -3.0$\pm$0.3 & 4760$\pm$226 & 0.7$\pm$0.5 \\ 
 331085117451894528 &  &  & True & 33.4272917 & 36.0950281 & 2457688.77 & 10.68 & -385.5$\pm$0.9 & -3.2$\pm$0.4 & 4684$\pm$255 & 1.4$\pm$0.5 \\ 
 331085117451894528 &  &  & True & 33.4272917 & 36.0950281 & 2457711.69 & 6.07 & -384.4$\pm$1.7 & -3.2$\pm$0.6 & 4772$\pm$453 & 0.9$\pm$0.7 \\ 
 331085117451894528 &  &  & True & 33.4272917 & 36.0950281 & 2457685.74 & 4.74 & -383.4$\pm$1.7 & -3.0$\pm$0.7 & 4783$\pm$495 & 1.7$\pm$0.9 \\ 
 331086526201161088 & 65 & 46 & True & 33.33975 & 36.1659431 & 2457374.8 & 8.84 & -400.4$\pm$0.7 & -2.5$\pm$0.3 & 4823$\pm$159 & 2.0$\pm$0.4 \\ 
 331086526201161088 & 65 & 46 & True & 33.33975 & 36.1659431 & 2457683.72 & 6.78 & -405.9$\pm$0.8 & -2.9$\pm$0.3 & 4535$\pm$215 & 0.9$\pm$0.4 \\ 
 331086526201161088 & 65 & 46 & True & 33.33975 & 36.1659431 & 2457688.77 & 5.08 & -405.1$\pm$1.0 & -2.0$\pm$0.7 & 5416$\pm$588 & 2.0$\pm$0.7 \\ 
 331086526201161088 & 65 & 46 & True & 33.33975 & 36.1659431 & 2457711.69 & 2.88 & -386.4$\pm$2.1 & -0.8$\pm$0.9 & 6811$\pm$855 & 2.2$\pm$1.0 \\ 
 331089446778920576 & 106 & 40 & True & 33.3189583 & 36.17939 & 2457374.8 & 27.68 & -382.8$\pm$0.6 & -3.5$\pm$0.1 & 4367$\pm$87 & 0.9$\pm$0.2 \\ 
 331089446778920576 & 106 & 40 & True & 33.3189583 & 36.17939 & 2457683.72 & 21.62 & -381.8$\pm$0.5 & -3.3$\pm$0.2 & 4507$\pm$95 & 1.0$\pm$0.3 \\ 
 331089446778920576 & 106 & 40 & True & 33.3189583 & 36.17939 & 2457688.77 & 23.86 & -381.9$\pm$0.5 & -3.3$\pm$0.1 & 4505$\pm$82 & 0.8$\pm$0.2 \\ 
 331089446778920576 & 106 & 40 & True & 33.3189583 & 36.17939 & 2457711.69 & 13.11 & -381.9$\pm$0.7 & -3.6$\pm$0.2 & 4253$\pm$135 & 0.7$\pm$0.3 \\ 
 331089446778920576 & 106 & 40 & True & 33.3189583 & 36.17939 & 2457685.74 & 12.85 & -382.2$\pm$0.8 & -3.2$\pm$0.3 & 4597$\pm$195 & 1.1$\pm$0.5 \\
 \hline
\end{tabular}
\caption{Individual observations of Tri~II members taken with MMT Hectochelle. This table is a subset of the full MMT catalogue.}
\label{tab:obs_MMT}
\end{sidewaystable}

\begin{table*}
    \centering
    \begin{tabular}{p{0.05\linewidth}| p{0.05\linewidth} |p{0.12\linewidth}|p{0.12\linewidth}|p{0.12\linewidth}|p{0.1\linewidth}}
     KCS2015 & MIC2016 & RA (ICRS) & Dec (ICRS) & HJD (days)  & $v_{\rm los} (\rm km~s^{-1})$\\ \hline
N/A & 8  & 33.2591667 & 36.2090836 & 2457284.07 &  $ -388.4 \pm 7.7 $\\
128 & N/A  & 33.3093333 & 36.1641944 & 2457303.10 &  $ -386.3 \pm 3.2 $\\
- & - & - & - & 2457416.70 &  $ -385.4 \pm 2.1 $\\
- & - & - & - & 2457416.80 &  $ -384.2 \pm 2.0 $\\
- & - & - & - & 2457639.10 &  $ -383.8 \pm 2.1 $\\
N/A & 31  & 33.4694167 & 36.2233611 & 2457370.70 &  $ -377.6 \pm 1.8 $\\
- & - & - & - & 2457416.70 &  $ -381.5 \pm 2.6 $\\
- & - & - & - & 2457639.10 &  $ -377.0 \pm 2.8 $\\
- & - & - & - & 2457284.07 &  $ -377.1 \pm 3.1 $\\
N/A & 29  & 33.3789583 & 36.1988889 & 2457370.70 &  $ -387.5 \pm 4.7 $\\
- & - & - & - & 2457284.07 &  $ -398.4 \pm 7.8 $\\
N/A & 27  & 33.3389583 & 36.1414167 & 2457370.70 &  $ -373.5 \pm 5.2 $\\
- & - & - & - & 2457416.70 &  $ -389.9 \pm 8.3 $\\
- & - & - & - & 2457284.07 &  $ -402.7 \pm 6.6 $\\
N/A & 26  & 33.3534583 & 36.1727222 & 2457370.70 &  $ -376.9 \pm 11.2 $\\
N/A & 24  & 33.3416667 & 36.1738611 & 2457416.70 &  $ -371.8 \pm 17.1 $\\
- & - & - & - & 2457284.07 &  $ -384.4 \pm 4.9 $\\
76 & 23  & 33.3358750 & 36.1629167 & 2457303.10 &  $ -391.0 \pm 3.0 $\\
- & - & - & - & 2457370.70 &  $ -384.6 \pm 3.1 $\\
- & - & - & - & 2457639.10 &  $ -384.0 \pm 3.0 $\\
- & - & - & - & 2457284.07 &  $ -389.2 \pm 3.6 $\\
N/A & 22  & 33.3028750 & 36.1470556 & 2457370.70 &  $ -381.5 \pm 3.4 $\\
- & - & - & - & 2457416.70 &  $ -381.4 \pm 3.3 $\\
- & - & - & - & 2457416.80 &  $ -384.9 \pm 3.7 $\\
- & - & - & - & 2457569.10 &  $ -381.1 \pm 23.5 $\\
- & - & - & - & 2457639.10 &  $ -376.6 \pm 6.6 $\\
- & - & - & - & 2457284.07 &  $ -388.3 \pm 3.8 $\\
116 & 21  & 33.3165000 & 36.1710556 & 2457303.10 &  $ -378.9 \pm 3.7 $\\
- & - & - & - & 2457370.70 &  $ -383.3 \pm 2.3 $\\
- & - & - & - & 2457416.70 &  $ -387.0 \pm 3.2 $\\
- & - & - & - & 2457416.80 &  $ -382.5 \pm 3.6 $\\
- & - & - & - & 2457639.10 &  $ -380.6 \pm 3.1 $\\
- & - & - & - & 2457284.07 &  $ -384.1 \pm 3.1 $\\
91 & 20  & 33.3305000 & 36.1925833 & 2457303.10 &  $ -387.3 \pm 3.1 $\\
- & - & - & - & 2457370.70 &  $ -379.1 \pm 1.9 $\\
- & - & - & - & 2457569.10 &  $ -388.2 \pm 3.9 $\\
- & - & - & - & 2457639.10 &  $ -379.2 \pm 2.4 $\\
- & - & - & - & 2457284.07 &  $ -380.0 \pm 2.9 $\\
N/A & 9  & 33.3639183 & 36.2251930 & 2457416.70 &  $ -388.9 \pm 7.7 $\\
- & - & - & - & 2457284.07 &  $ -406.0 \pm 5.1 $\\
106 & 40  & 33.3189583 & 36.1793889 & 2457303.10 &  $ -383.6 \pm 1.5 $\\
- & - & - & - & 2457416.70 &  $ -382.1 \pm 1.5 $\\
- & - & - & - & 2457416.80 &  $ -383.3 \pm 1.5 $\\
- & - & - & - & 2457569.10 &  $ -382.8 \pm 1.5 $\\
- & - & - & - & 2457284.07 &  $ -380.5 \pm 2.3 $\\
65 & 46  & 33.3397500 & 36.1659444 & 2457303.10 &  $ -375.8 \pm 1.7 $\\
- & - & - & - & 2457416.70 &  $ -386.7 \pm 1.6 $\\
- & - & - & - & 2457569.10 &  $ -376.3 \pm 4.0 $\\
- & - & - & - & 2457639.10 &  $ -384.6 \pm 1.6 $\\
- & - & - & - & 2457284.07 &  $ -373.8 \pm 2.4 $\\
\hline 
\end{tabular}
\caption{Individual observations of Tri~II members from Keck DEIMOS used in this analysis after adding a $\delta_v = -1.33 \rm~km~s^{-1}$ zero-point offset correction. The first epochs for each star are listed with the star's ID numberings and sky position while any additional epochs for the same star are in subsequent rows with "-" in the ID and sky position columns.}
\label{tab:obs_keck}
\end{table*}


\bsp	
\label{lastpage}
\end{document}